\documentclass[prd,reprint]{revtex4-2}

\usepackage[english]{babel}
\usepackage{graphicx}
\usepackage{amsmath}
\usepackage{amsfonts}
\usepackage[utf8]{inputenc}
\usepackage{xcolor}
\usepackage{physics}
\usepackage{overpic,color}
\usepackage{hyperref}
\usepackage{dsfont}
\usepackage[charter,expert]{mathdesign}
\usepackage{enumitem}
\usepackage{pgfplots}
\usepackage{tikz}
\usetikzlibrary{calc}
\usetikzlibrary{arrows.meta, arrows}
\usetikzlibrary{decorations.pathmorphing}
\tikzset{snakeit/.style={decorate, decoration=snake}}

\newcommand{\hamiltonianglobal}{\hat{\mathsf{H}}_{\text{tot}}}
\newcommand{\hamiltonianglobalfree}{\hat{\mathsf{H}}_{\text{tot},0}}
\newcommand{\hamiltoniansystem}{\hat{H}}
\newcommand{\hamiltonianclock}{\hat{H}_{\text{C}}}

\newcommand{\energyglobal}{E}
\newcommand{\energysystem}{\varepsilon}
\newcommand{\energyclock}{\mathcal{E}}

\newcommand{\interactionfunction}{W}
\newcommand{\interaction}{\hat{\textsf{\interactionfunction}}}
\newcommand{\potentialfunction}{V}
\newcommand{\potentialsystem}{\hat{\potentialfunction}}

\newcommand{\couplingstrength}{g}

\newcommand{\stateglobal}{\Psi}
\newcommand{\statesystem}{\varphi}

\newcommand{\stateclock}{\chi}

\newcommand{\prm}{t}
\newcommand{\prmalt}{s}

\newcommand{\hilbert}{\mathcal{H}}
\newcommand{\hilbertglobal}{\hilbert_{\textrm{tot}}}
\newcommand{\hilbertsystem}{\hilbert}
\newcommand{\hilbertclock}{\hilbert_{\text{C}}}

\newcommand{\ketglobal}[1]{\ket*{#1}\!\rangle}
\newcommand{\ketsystem}[1]{\ket*{#1}}

\newcommand{\ketclock}[1]{\ket{#1}_{\text{C}}}

\newcommand{\braglobal}[1]{\bra*{\!\langle#1}}
\newcommand{\brasystem}[1]{\bra*{#1}}

\newcommand{\braketsystem}[2]{\braket{#1}{#2}}

\newcommand{\braketclockglobal}[2]{\braket*{#1}{#2}\!\rangle_{\text{C}}}

\newcommand{\melglobal}[3]{\mel*{\!\langle #1}{#2}{#3 \rangle\!}}
\newcommand{\melclock}[3]{\mel{#1}{#2}{#3}_{\text{C}}}
\newcommand{\melclockglobal}[3]{\mel*{#1}{#2}{#3}\!\rangle_{\text{C}}}

\newcommand{\ident}{\hat{\mathds{1}}}
\newcommand{\identsystem}{\ident}
\newcommand{\identclock}{\ident_{\text{C}}}

\newcommand{\dyadglobal}[2]{\ketglobal{#1} \hspace{-0.93mm} \braglobal{#2}}
\newcommand{\dyadclock}[2]{\dyad{#1}{#2}_{\text{C}}}

\newcommand{\dyadsystemop}[2]{\dyad{#1}{#2}_{\text{S}}}

\newcommand{\projectorglobal}{\hat{\textsf{P}}_{\stateglobal}}

\newcommand{\projectorclock}{\hat{\textsf{P}}_{\stateclock}}
\newcommand{\projectorclockdyn}{\hat{\textsf{P}}_{\stateclock(\prm)}}
\newcommand{\projectorclockinit}{\hat{\textsf{P}}_{\stateclock_0}}

\newcommand{\normalization}{N}

\newcommand{\unitaryglobal}{\hat{\mathsf{U}}_0}
\newcommand{\unitaryglobaldag}{\unitaryglobal^{\dagger}}

\newcommand{\unitarysysalt}[1]{e^{-i#1\hamiltoniansystem_0}}
\newcommand{\unitarysysaltdag}[1]{e^{i#1\hamiltoniansystem_0}}
\newcommand{\unitaryclock}{\hat{U}_{\text{C}}}

\newcommand{\projectordegenspace}{\hat{\mathsf{P}}_0}
\newcommand{\projectordegenspacecomp}{\hat{\mathsf{Q}}_0}
\newcommand{\projectordegenspacewozero}{\hat{\overline{\mathsf{P}}}_0}

\newcommand{\sigmax}{\hat{\sigma}_x}
\newcommand{\sigmay}{\hat{\sigma}_y}
\newcommand{\sigmaz}{\hat{\sigma}_z}
\newcommand{\sigmai}{\hat{\sigma}_i}

\newcommand{\angmomz}{\hat{J}_z}
\newcommand{\angmomi}{\hat{J}_i}

\allowdisplaybreaks

\begin{document}

\title{A link between static and dynamical perturbation theory}
\author{Sebastian Gemsheim}
\email{sebgem@protonmail.com}
\affiliation{Max Planck Institute for the Physics of Complex Systems, N\"othnitzer Stra\ss e 38, 01187 Dresden, Germany}
\affiliation{Institute for Photonic Quantum Systems (PhoQS), Warburger Stra\ss e 100, 33098 Paderborn, Germany}

\begin{abstract}
Dynamics, the physical change in time and a pillar of natural sciences, can be regarded as an emergent phenomenon when the system of interest is part of a larger, static one. This ``relational approach to time'', in which the system's environment provides a temporal reference, does not only provide insight into foundational issues of physics, but holds the potential for a deeper theoretical understanding as it intimately links statics and dynamics. Reinforcing the significance of this connection, we demonstrate, based on recent progress [Phys.~Rev.~Lett.~\textbf{131}, 140202 (2023)], the role of emergent time as a vital link between time-\emph{independent} and time-\emph{dependent} perturbation theory in quantum mechanics. We calculate first order contributions, which are often the most significant, and discuss the issue of degenerate spectra. Based on our results, we envision future applications for the calculation of dynamical phenomena based on a single pure energy eigenstate.
\end{abstract}

\maketitle

%%%%%%%%%%%%%%%%%%%%%%%%%%%%%%%%%%%%%%%%%%%%%%%%%%%%
%%% BODY

\section{Introduction}

The effectiveness of mathematical equations to describe the astonishing complexity of very distinct phenomena in the physical world is simply extraordinary~\cite{Wigner1960}. Unfortunately, and more often than not, these equations escape a closed analytical treatment in many cases. While challenging, this fact does not preclude at least a partial understanding of their physical consequences. Perturbation theory allows scientists to deduce valuable insight even in the face of insolubility, by sacrificing quantitative accuracy. This indispensable tool provides analytical solutions in the vicinity of known unperturbed problems. If the perturbation is sufficiently weak, then a finite number of expansion terms in powers of its strength may adequately approximate the solution and offers qualitative understanding of the original problem. 

Already Schr{\"o}dinger, in his famous papers~\cite{Schroedinger1926_3, Schroedinger1926_4} on quantum mechanics, employed schemes that are still in use today. Based on the expansion in terms of a small coupling parameter, these methods are prominently categorized by two classes: time-\emph{independent} and time-\emph{dependent}. The former is used to compute the energetic shifts and modifications of energy eigenstates due to slight changes of the original Hamiltonian. In contrast, the latter focuses on dynamical changes of an arbitrary initial state when its original (often time-independent) Hamiltonian is perturbed by a time-dependent term. Although they rely on the same principle, both schemes are distinct due to their domains of application being in the static and dynamical realm, respectively. 

Bridging both regimes has often offered interesting new ways to infer static properties from dynamical considerations and vice versa. Particular examples are periodically driven systems~\cite{Sambe1973}, the Fourier transformation of autocorrelation functions for the spectral analysis of quantum states~\cite{Grossmann2018}, spectral form factors~\cite{delCampo2017}, eigenstate thermalization hypothesis~\cite{Deutsch1991, Srednicki1994, Deutsch2018} and ergodicity breaking in many-body systems~\cite{Abanin2019, Serbyn2021, Moudgalya2022} to name a few. Moreover, this connection even encompasses foundational issues such as the origin of time.

The non-existence of a genuine time operator~\cite{Pegg1998, Muga2008, Dodonov2015} and the appearance of time as an external parameter in quantum mechanics has puzzled many generations of physicists~\cite{Hilgevoord2002, Hilgevoord2005}. An intriguing approach to reconcile this discrepancy in comparison with other quantum mechanical operators like position was put forth by Page and Wootters in 1983~\cite{Page1983}. They postulated a static state of the universe in which time emerges only for a subsystem of the whole in relation to the state of its complement, the rest of the universe. Thus, in the relational approach to time, dynamics appears from a timeless state by means of a splitting of the whole into parts. The complement (C) of a primary system (short ``system'' S) is often denoted a ``clock'' due to its function as a temporal reference for S. In general, the separation into subsystems entails an interaction term in the global Hamiltonian, but the original work~\cite{Page1983} did not consider such almost inevitable couplings of a system with its complement. Most often these scenarios with interactions have been dealt with by means of a semiclassical approximation~\cite{Briggs2000, Briggs2001, Briggs2007, Braun2004, Briggs2015, Schild2018} in which the clock states are taken as Wentzel–Kramers–Brillouin wave functions where the phase is dominated by a classical action. Recent progress~\cite{Gemsheim2023a, Gemsheim2023phd} established how this setting can be treated in the most general case for arbitrary Hamiltonians. An important feature of the interacting case is the appearance of an effective potential for the system which derives from the global static coupling between system and clock. This additional term becomes implicitly time-dependent through its explicit dependence on the state of the clock. 

The main goal of our paper is to reveal, in full generality, the role of "relational time" as the central element for deriving within quantum mechanical perturbation theory the dynamical parts of a subsystem from the purely static ones of the global system. After a brief summary of time-dependent perturbation theory (TDPT) and of the main findings in Ref.~\cite{Gemsheim2023a}, we demonstrate how relational time provides a direct link between the former and time-independent perturbation theory (TIPT) through the aforementioned effective system potential. Following this, we discuss the consequences of degeneracy for our results and provide an instructive example. Starting points for successive research are given at the end after a short conclusion.

%%%%%%%%%%%%%%%%%%%%%%%%%%%%%%%%%%%%%%%%%%%%%%%%%%%%

\section{Prerequisites}

\subsection{Time-dependent perturbation theory}

The dynamics of a quantum mechanical system of interest, characterized by a time-independent Hamiltonian $\hamiltoniansystem_0$, is completely determined at all times once the eigenvalues and eigenstates of $\hamiltoniansystem_0$ are known. Being fully isolated, such a free time evolution is certainly of interest, but eludes any probing from the outside. Bringing the system into contact with its environment through a time-dependent potential $\couplingstrength \potentialsystem(\prm)$ makes it possible to alter the evolution of an initial system state $\ketsystem{\statesystem(0)} \in \hilbertsystem$. Its change in Hilbert space $\hilbertsystem$ is governed by the time-dependent Schr{\"o}dinger equation (TDSE)
\begin{equation}
    i \dv{\prm} \ketsystem{\statesystem(\prm)} 
    = \hamiltoniansystem(\prm) \ketsystem{\statesystem(\prm)}
    \label{eq:tdse}
\end{equation}
with the combined Hamiltonian
\begin{equation}
    \hamiltoniansystem(\prm) 
    = \hamiltoniansystem_0 + \couplingstrength \potentialsystem(\prm) \,.
\end{equation}
We introduce a dimensionless coupling strength $\couplingstrength$ and use the atomic unit system, unless stated otherwise. Analytical solutions to the TDSE are exceedingly rare. Nevertheless, perturbation theory in orders of $\couplingstrength$ allows for qualitative and even quantitative expressions for $\ketsystem{\statesystem(\prm)}$. 
Although mathematical issues can arise in perturbation theory~\cite{Bender1999, Picasso2022}, we restrict ourselves to first order perturbations for $\couplingstrength \ll 1$ with the aim of keeping the mathematical analysis simple and elucidating the general principle of our findings. 
In this case, the state reads
\begin{equation}
    \ketsystem{\statesystem(\prm)}
     \approx \unitarysysalt{\prm}  \left( \identsystem -i \couplingstrength \int_0^{\prm} \dd{\prmalt} \,\unitarysysaltdag{\prmalt}  \potentialsystem(\prmalt) \, \unitarysysalt{\prmalt}   \right) \ketsystem{\statesystem(0)} 
     \label{eq:first_order_tdpt}
\end{equation}
with the lowest order correction, which frequently constitute the most significant element of perturbation theory, as, for example, in Fermi's Golden Rule~\cite{Ballentine2014}. As the magnitude of the perturbation is roughly determined by the strength and the temporal extent of the external potential, Eq.~\eqref{eq:first_order_tdpt} adequately approximates the actual evolution for sufficiently short times $\prm$, even for $g=1$.

%%%%%%%%%%%%%%%%%%%%%%%%%%%%%%%%%%%%%%%%%%%%%%%%%%%%%%%%%%%%%%%%%%%%%%%%%%%%%%%%

\subsection{Relational time}

Any external potential will necessarily depend on the state of the environment surrounding the system. However, the way in which such a $\potentialsystem(\prm)$ arises may not be unique. In fact, its appearance can even be tightly linked to the quantum mechanical origin of the parameter $\prm$ itself. 
Page and Wootters~\cite{Page1983, Wootters1984} seminally advocated for the conceptual shift in perspective of time manifesting as the relation of a system to its environment. 
But if time is only meaningful for a subsystem, then the global whole, in which the system is embedded, must be truly static. Such a "timeless" global state $\ketglobal{\stateglobal}$ solves the 
time-independent Schr\"odinger equation (TISE)
\begin{equation}
    (\hamiltonianglobal - \energyglobal) \ketglobal{\stateglobal} 
    = 0 
    \label{eq:tise}
\end{equation}
for the total Hamiltonian $\hamiltonianglobal$ and energy $\energyglobal$. 
Splitting off a (small) system introduces the product $\hilbertglobal = \hilbertsystem \otimes \hilbertclock$, where the complement C comprises the environment.
The double angle brackets serve as a reminder of this bipartite structure and designate elements of the global Hilbert space $\hilbertglobal$. 
A generic total Hamiltonian $\hamiltonianglobal$ has the form
\begin{equation}
    \hamiltonianglobal 
    = \hamiltonianglobalfree + \couplingstrength \interaction
\end{equation}
under such a partitioning 
with the free subsystem Hamiltonians $\hamiltonianglobalfree = \hamiltoniansystem_0 \otimes \identclock + \identsystem \otimes \hamiltonianclock$ and an interaction term $\couplingstrength\interaction$ between system and environment. 

Since the system is embedded, its state $\ketsystem{\statesystem}$ derives from the partial projection~\footnote{We neglect identity operations on subsystems, unless it is useful to state them explicitly.}
\begin{equation}
    \ketsystem{\statesystem} 
    = \frac{1}{\sqrt{\normalization}} \braketclockglobal{\stateclock}{\stateglobal}
    \label{eq:relational_system_state}
\end{equation}
of an environment state $\ketclock{\stateclock} \in \hilbertclock$ onto $\stateglobal$. Simply put, the system is uniquely related to (or conditioned on) $\stateclock$ through the quantum correlations contained in the global state. The normalization factor $\normalization = \melglobal{\stateglobal}{\projectorclock}{\stateglobal}$ with the clock projector $\projectorclock = \dyadclock{\stateclock}{\stateclock}$ ensures $\braketsystem{\statesystem}{\statesystem} = 1$.
To understand how a time parameter emerges in this "relational approach", we invoke the invariance
\begin{equation}
    e^{i\prm(\hamiltonianglobal - \energyglobal)} \ketglobal{\stateglobal} 
    = \ketglobal{\stateglobal} \,,
    \label{eq:invariance}
\end{equation}
which further highlights the absence of any change on the global level.
Differentiation with respect to the real symmetry parameter $\prm$ shows that Eq.~\eqref{eq:invariance} is equivalent to the TISE~\eqref{eq:tise}. 
As shown in Ref.~\cite{Gemsheim2023a}, Eqs.~\eqref{eq:invariance} and~\eqref{eq:relational_system_state} give rise to the system TDSE
\begin{equation}
    i \dv{\prm} \ketsystem{\statesystem(\prm)} = \left[ \hamiltoniansystem_0 + \couplingstrength \potentialsystem(\prm) \right] \ketsystem{\statesystem(\prm)} \,,
\end{equation}
which involves the clock states $\ketclock{\stateclock(\prm)} = \unitaryclock(\prm) \ketclock{\stateclock_0} = e^{-i\prm (\hamiltonianclock - \energyglobal)} \ketclock{\stateclock_0}$ with initial $\ketclock{\stateclock_0}$. 
Thus, the environment acts as a temporal reference (or simply a "clock") for the system. Surprisingly, this leads to the concurrent emergence of time and an effective system potential
\begin{equation}
    \couplingstrength \potentialsystem(\prm) 
    = \frac{\couplingstrength}{\normalization(\prm)} \left[ \melclock{\stateclock(\prm)}{ \bigl\{ \interaction, \projectorglobal \bigr\} }{\stateclock(\prm)} - \Re \melglobal{\stateglobal}{\interaction\projectorclockdyn}{\stateglobal} \right] \,,
    \label{eq:effective_system_potential}
\end{equation}
depending directly on $\stateclock$, through which its dynamical character is imprinted. Here, $\{\hat{A}, \hat{B}\} = \hat{A}\hat{B} + \hat{B}\hat{A}$ denotes the anti-commutator.
We define the global projector $\projectorglobal = \dyadglobal{\stateglobal}{\stateglobal}$ and note the inconsequential use of slightly different definitions than in Ref.~\cite{Gemsheim2023a}. Crucially, all system dynamics is encoded through the entanglement in a single energy eigenstate, rendering quantum correlations in $\ketglobal{\stateglobal}$ a prerequisite for time that is automatically fulfilled for generic $\interaction$'s. 

Although the parameter $\prm$ cannot strictly be associated with time~\cite{Gemsheim2023a}, as it is just a symmetry parameter, we nevertheless use the symbol $t$ and assume that a simple linear relationship between this parametrization and a physical clock property exists. The latter provides a physical time for the system as their relation is invariant under reparametrizations of $\prm$. 

Relational dynamics features an additional distinct property. In general, the initial relational state and the effective potential inherently depend on the global state. Any change in the total state, or the coupling strength later on, thus causes a simultaneous change in both. Such a behavior may not always be desired for practical applications. Finding ways to circumvent this concurrent change is however part of ongoing research and cannot be addressed in this work. \\

As mentioned in the Introduction, we aim to establish a link between the time-dependent perturbed state~\eqref{eq:first_order_tdpt} and static global states with time-independent corrections. 
To this end, the dynamics of the system state~\eqref{eq:relational_system_state} is investigated based on the simultaneous corrections in the global energy $\energyglobal$ and $\ketglobal{\stateglobal}$ due to coupling between system and environment. The explicit linear appearance of $\couplingstrength$ in the effective potential~\eqref{eq:effective_system_potential} anticipates the possibility to use the unperturbed global state $\ketglobal{\stateglobal^{(0)}}$ for its calculation, while still being correct in first order of $\couplingstrength$. Following a brief recap of TIPT in quantum mechanics, we show the correctness of this qualitative analysis.

%%%%%%%%%%%%%%%%%%%%%%%%%%%%%%%%%%%%%%%%%%%%%%%%%%%%%%%%%%%%%%%%%%%%%%%%%%%%%%%%

\subsection{Time-independent perturbation theory}

Without an explicit derivation, we simply state the result of TIPT on the global level as they are in the standard repertoire of each physics curriculum. An expansion of the TISE~\eqref{eq:tise} in orders of $\couplingstrength$ yields the energy
\begin{equation}
    \energyglobal 
    = \energyglobal^{(0)} + \couplingstrength\energyglobal^{(1)} + \mathcal{O}(\couplingstrength^2) \,.
    \label{eq:global_energy_tipt}
\end{equation}
with the leading order term
\begin{equation}
    \energyglobal^{(1)} = \melglobal{\stateglobal^{(0)}}{\interaction}{\stateglobal^{(0)}} \,.
    \label{eq:global_energy_correction_first_order}
\end{equation}
The eigenstates of the uncoupled Hamiltonian $\hamiltonianglobalfree$ with unperturbed eigenenergy $\energyglobal^{(0)}$, i.e., $( \hamiltonianglobalfree - \energyglobal^{(0)} ) \ketglobal{\stateglobal^{(0)}} = 0$, provide the zeroth order term in the global state
\begin{equation}
    \ketglobal{\stateglobal}
    = \ketglobal{\stateglobal^{(0)}} + \couplingstrength \ketglobal{\stateglobal^{(1)}} + \mathcal{O}(\couplingstrength^2) \,.
    \label{eq:global_state_tipt}
\end{equation}
Its correction for sufficiently weak couplings fulfills the relation
\begin{equation}
    \Bigl[ \energyglobal^{(0)} - \hamiltonianglobalfree \Bigr] \ketglobal{\stateglobal^{(1)}} 
    = \Bigl[ \interaction - \energyglobal^{(1)}\Bigr] \ketglobal{\stateglobal^{(0)}} \,.
    \label{eq:relation_global_state_correction}
\end{equation}
For now, we assume the unperturbed energy level $\energyglobal^{(0)}$ to be strictly non-degenerate. Although we do not need it for the following derivation, this allows us to provide the formal solution 
\begin{equation}
    \ketglobal{\stateglobal^{(1)}} 
    = \Bigl( \ident - \projectorglobal^{(0)} \Bigr) \Bigl( \energyglobal^{(0)} - \hamiltonianglobalfree \Bigr)^{-1} \Bigl( \ident - \projectorglobal^{(0)} \Bigr) \interaction \ketglobal{\stateglobal^{(0)}} 
    \label{eq:global_state_correction}
\end{equation}
with $\projectorglobal^{(0)} = \dyadglobal{\stateglobal^{(0)}}{\stateglobal^{(0)}}$
for the global state correction~\cite{Sakurai2017}. A well-known fact of perturbation theory is that such an approach can only be accurately used in first order if the perturbation facilitated by $\interaction$ is small compared to the energy splittings of the unperturbed problem~\cite{Picasso2022}. 

%%%%%%%%%%%%%%%%%%%%%%%%%%%%%%%%%%%%%%%%%%%%%%%%%%%%

\section{Relational state of the system for weak coupling}

For the expansion of the relational system state~\eqref{eq:relational_system_state} in
$\couplingstrength$, we omit expressions for terms higher than first order due to the weak coupling assumption. Our derivation starts with an analysis of the initial state
\begin{equation}
    \ketsystem{\statesystem(0)} 
    = \frac{\braketclockglobal{\stateclock_0}{\stateglobal}}{\sqrt{\normalization(0)}} 
    \approx \frac{\braketclockglobal{\stateclock_0}{\stateglobal^{(0)}} + \couplingstrength \braketclockglobal{\stateclock_0}{\stateglobal^{(1)}} }{\sqrt{\normalization(0)}} 
\end{equation}
where the normalization factor reads $\normalization(0) \approx \melglobal{\stateglobal^{(0)}}{\projectorclockinit}{\stateglobal^{(0)}} + 2 \couplingstrength \Re \melglobal{\stateglobal^{(0)}}{\projectorclockinit}{\stateglobal^{(1)}}$. A comparison to Eq.~\eqref{eq:first_order_tdpt} suggests the definition of the dynamical zeroth order as 
\begin{equation}
    \ketsystem{\statesystem^{(0)}(\prm)} 
    = \unitarysysalt{\prm} \ketsystem{\statesystem(0)} \,.
    \label{eq:system_state_zeroth_order_relational}
\end{equation}
It may seem peculiar that the lowest order contribution to dynamics already includes a static correction proportional to $\couplingstrength$. This behavior depends on the intrinsic and simultaneous dependence of the effective potential and the initial state on the global state. Therefore, we separate the change in $\ketsystem{\statesystem(0)}$ from the occurrence of $\potentialsystem$ as the focus lies on dynamical perturbations. 

In the following, we want to decompose the time-evolved system state~\eqref{eq:relational_system_state} including the energy correction~\eqref{eq:global_energy_tipt}, i.e., $\energyglobal \approx \energyglobal^{(0)} + \couplingstrength\energyglobal^{(1)}$. Knowing the zeroth order form~\eqref{eq:system_state_zeroth_order_relational}, we find, up to first order in $\couplingstrength$, that
\begin{subequations}
\begin{align}
    \ketsystem{\statesystem(\prm)} 
    &\approx \frac{1}{\sqrt{\normalization(\prm)}} \left[ \braketclockglobal{\stateclock(\prm)}{\stateglobal^{(0)}} + \couplingstrength \braketclockglobal{\stateclock(\prm)}{\stateglobal^{(1)}} \right] \\
    &= \frac{e^{-i\prm\couplingstrength\energyglobal^{(1)}}}{\sqrt{\normalization(\prm)}} \Biggl[ \unitarysysalt{\prm} \left( \braketclockglobal{\stateclock_0}{\stateglobal^{(0)}} + \couplingstrength \braketclockglobal{\stateclock_0}{\stateglobal^{(1)}}  \right)  \nonumber\\
    &\quad + \couplingstrength \Bigl(  \braketclockglobal{\stateclock(\prm)}{\stateglobal^{(1)}} - \unitarysysalt{\prm} \braketclockglobal{\stateclock_0}{\stateglobal^{(1)}} \Bigr) \Biggr] \\
    &= e^{-i\prm\couplingstrength\energyglobal^{(1)}} \unitarysysalt{\prm}  \Biggl[ \sqrt{\frac{\normalization(0)}{\normalization(\prm)}} \ketsystem{\statesystem(0)} \nonumber\\
    &\quad + \frac{\couplingstrength}{\sqrt{\normalization(\prm)}} \melclockglobal{\stateclock_0}{ \Bigl( \unitaryglobaldag(\prm) - \ident \Bigr) }{\stateglobal^{(1)}} \Biggr] 
    \label{eq:relational_state_intermediate}
\end{align}
\end{subequations}
with the global unitary operator $\unitaryglobaldag(\prm) = \exp[-i\prm(\energyglobal^{(0)}-\hamiltonianglobalfree) ]$. 

As a first step toward the desired form of the evolution, we partially Taylor expand
\begin{equation}
    \sqrt{\frac{\normalization(0)}{\normalization(\prm)}} 
    \approx 1 - \frac{\normalization(\prm) - \normalization(0)}{2\normalization(0)} \,.
    \label{eq:normalization_expansion}
\end{equation}
by using the fact that the difference
\begin{equation}
    \normalization(\prm) - \normalization(0)
    \approx 2 \couplingstrength \Re \melglobal{\stateglobal^{(0)}}{\Bigl( \projectorclockdyn - \projectorclockinit \Bigr) }{\stateglobal^{(1)}} 
\end{equation}
scales linearly in the coupling strength in leading order. 
Expanding the scalar exponential factor in front of Eq.~\eqref{eq:relational_state_intermediate} to first order comprises the second step. In combination with the previous equation, we yield
\begin{align}
    e^{-i\prm\couplingstrength\energyglobal^{(1)}} \sqrt{\frac{\normalization(0)}{\normalization(\prm)}} \approx 1 - i\prm\couplingstrength\energyglobal^{(1)} - \frac{\normalization(\prm) - \normalization(0)}{2\normalization(0)} 
\end{align}
and take $\couplingstrength e^{-i\prm\couplingstrength\energyglobal^{(1)}} \approx \couplingstrength$ for the second term in squared brackets in Eq.~\eqref{eq:relational_state_intermediate}.
The third step consists of a further evaluation of the correction term. To this end, the operator relation
\begin{equation}
    \Bigl( e^{-i\hat{A}\prm} - \ident \Bigr) 
    = \int_0^{\prm} \dd{\prmalt} \dv{\prmalt} e^{-i\hat{A}\prmalt} 
    = - i \int_0^{\prm} \dd{\prmalt} e^{-i\hat{A}\prmalt} \hat{A}
\end{equation}
allows us to express
\begin{subequations}
\begin{align}
    &\melclockglobal{\stateclock_0}{ \Bigl( \unitaryglobaldag(\prm) - \ident \Bigr) }{\stateglobal^{(1)}} \nonumber\\
    &\qquad = -i \int_0^{\prm} \dd{\prmalt} \melclockglobal{\stateclock_0}{\unitaryglobaldag(\prmalt) \Bigl( \energyglobal^{(0)} - \hamiltonianglobalfree \Bigr) }{\stateglobal^{(1)}} \\
    \label{eq:energy_relation_first_order_correction_for_time_integral}
     &\qquad = -i \int_0^{\prm} \dd{\prmalt} \melclockglobal{\stateclock_0}{\unitaryglobaldag(\prmalt) \Bigl( \interaction - \energyglobal^{(1)} \Bigr) }{\stateglobal^{(0)}} \\
    &\qquad = -i \int_0^{\prm} \dd{\prmalt} \unitarysysaltdag{\prmalt} \melclockglobal{\stateclock(\prmalt)}{\interaction }{\stateglobal^{(0)}} + i \prm \energyglobal^{(1)} \braketclockglobal{\stateclock_0}{\stateglobal^{(0)}}
\end{align}
\end{subequations}
with the help of $\unitaryglobaldag(\prmalt) \ketglobal{\stateglobal^{(0)}} = \ketglobal{\stateglobal^{(0)}}$ and Eq.~\eqref{eq:relation_global_state_correction}. Substituting the last line into the previous expression for the relational system state motivates the additional approximations $\couplingstrength / \sqrt{\normalization(\prm)} \approx \couplingstrength / \sqrt{\normalization(0)}$ and
\begin{equation}
    \frac{\couplingstrength}{ \sqrt{\normalization(\prm)} } \braketclockglobal{\stateclock_0}{\stateglobal^{(0)}} 
    \approx \couplingstrength \ketsystem{\statesystem(0)} \,.
    \label{eq:approximation_initial_state}
\end{equation}
As a result, we get
\begin{align}
    \ketsystem{\statesystem(\prm)} 
    &\approx \unitarysysalt{\prm} \Biggl[ \left( 1 - \frac{\normalization(\prm) - \normalization(0)}{2\normalization(0)} \right) \ketsystem{\statesystem(0)} \nonumber\\
    &\quad -i \frac{\couplingstrength}{\sqrt{\normalization(0)}} \int_0^{\prm} \dd{\prmalt} \unitarysysaltdag{\prmalt}  \melclockglobal{\stateclock(\prmalt)}{\interaction }{\stateglobal^{(0)}} \Biggr] \,.
\end{align}
The last term, featuring the partial projection onto the global coupling, represent a hurdle as it is not immediately clear how it can be transformed to a linear operator acting on the system state.
However, utilizing a key finding of Ref.~\cite{Gemsheim2023a}, we can decompose
\begin{align}
    &\melclockglobal{\stateclock}{ \interaction }{\stateglobal^{(0)}} \nonumber\\
    &\qquad = \left[ \potentialsystem^{(0)} - i \frac{ \Im \melglobal{\stateglobal^{(0)}}{ \interaction \projectorclock }{\stateglobal^{(0)}} }{ \melglobal{\stateglobal^{(0)}}{ \projectorclock }{\stateglobal^{(0)}} } \right] \unitarysysalt{\prmalt}  \braketclockglobal{\stateclock_0}{\stateglobal^{(0)}}
    \label{eq:expansion_effective_system_potential}
\end{align}
into an effective Hermitian potential
\begin{align}
    \potentialsystem^{(0)}
    &= \frac{ \melclock{\stateclock}{ \left\{ \interaction, \projectorglobal^{(0)} \right\} }{\stateclock} }{\melglobal{\stateglobal^{(0)}}{ \projectorclock }{\stateglobal^{(0)}}}  - \frac{ \Re \melglobal{\stateglobal^{(0)}}{ \interaction  \projectorclock }{\stateglobal^{(0)}} }{ \melglobal{\stateglobal^{(0)}}{ \projectorclock }{\stateglobal^{(0)}} } 
    \label{eq:effective_potential_zeroth_order}
\end{align}
and a purely imaginary scalar acting on the unnormalized system state $\unitarysysalt{\prmalt} \braketclockglobal{\stateclock_0}{\stateglobal^{(0)}}$. The imaginary scalar in Eq.~\eqref{eq:expansion_effective_system_potential} is proportional to the time derivative of the normalization factor in first order, i.e.,
\begin{subequations}
\begin{align}
    i \dv{\prm} \normalization(\prm)
    &= -2i \couplingstrength \Im \melglobal{\stateglobal}{ \interaction \projectorclock(\prm) }{\stateglobal} \\
    &\approx -2i \couplingstrength \Im \melglobal{\stateglobal^{(0)}}{ \interaction \projectorclock(\prm) }{\stateglobal^{(0)}} \,.
\end{align}
\end{subequations}
Thus, by means of $\couplingstrength / \melglobal{\stateglobal^{(0)}}{ \projectorclock }{\stateglobal^{(0)}}  \approx \couplingstrength / \normalization(0) $ and Eq.~\eqref{eq:approximation_initial_state} we obtain
\begin{align}
    &\frac{\couplingstrength}{\sqrt{\normalization(0)}} \unitarysysaltdag{\prmalt} \melclockglobal{\stateclock(\prmalt)}{ \interaction }{\stateglobal^{(0)}} \nonumber\\
    &\qquad \approx \Biggl[ \couplingstrength \unitarysysaltdag{\prmalt} \, \potentialsystem^{(0)}(\prmalt) \, \unitarysysalt{\prmalt} + \frac{i}{2\normalization(0)} \dv{\normalization(\prmalt)}{\prmalt} \Biggr] \ketsystem{\statesystem(0)}
    \label{eq:expansion_effective_system_potential_second}
\end{align}
Remarkably, the contribution from the imaginary scalar in Eq.~\eqref{eq:expansion_effective_system_potential} cancels exactly with the partial expansion of the normalization factor in Eq.~\eqref{eq:normalization_expansion}. The combination of all intermediate expressions implies our final result and main finding
\begin{equation}
    \ketsystem{\statesystem(\prm)} 
    \approx \unitarysysalt{\prm} \left[ \ident -i \couplingstrength  \int_0^{\prm} \dd{\prmalt} \unitarysysaltdag{\prmalt} \, \potentialsystem^{(0)}(\prmalt) \, \unitarysysalt{\prmalt} \right] \ketsystem{\statesystem(0)} \,.
    \label{eq:final_result_tdpt}
\end{equation}
in the form of Eq.~\eqref{eq:first_order_tdpt}. Therefore, we can answer the question about a link between TIPT and TDPT in the affirmative. While still being correct to first order, one can also substitute $\potentialsystem^{(0)}(\prmalt)$ by $\potentialsystem(\prmalt)$ if desired. The former may, however, be easier to compute.

%%%%%%%%%%%%%%%%%%%%%%%%%%%%%%%%%%%%%%%%%%%%%%%%%%%%%%%%%%%%%%%%%%%%%%%%%%%%%%%%

\section{Degeneracy}
While the previous treatment assumed a non-degenerate global energy spectrum, we now show that our results are valid and unchanged even in the presence of symmetries leading to degenerate energy subspaces for the unperturbed global Hamiltonian $\hamiltonianglobalfree$. For simplicity, we assume a degeneracy that is lifted at first order. In this case, we define a projector 
\begin{equation}
    \projectordegenspace
    = \sum_{m=1}^{D} \dyadglobal{\stateglobal^{(0)}_m}{\stateglobal^{(0)}_m}
\end{equation}
where $D$ is the degeneracy of the unperturbed energy level $\energyglobal^{(0)}$ with the corresponding eigenstates $\ketglobal{\stateglobal^{(0)}_m}$ spanning the degenerate subspace. In addition, we define its complement $\projectordegenspacecomp = \ident - \projectordegenspace$. As is widely known~\cite{Picasso2022}, a number of convenient projections onto the TISE leads to the relation
\begin{equation}
    \projectordegenspace \interaction \projectordegenspace \ketglobal{\stateglobal^{(0)}}
    = \energyglobal^{(1)} \ketglobal{\stateglobal^{(0)} } \,,
\end{equation}
providing a two-fold insight. First, the unperturbed zeroth order state must be chosen as an eigenstate of the reduced operator $\projectordegenspace \interaction \projectordegenspace$ in order to ensure a smooth dependence of the global state on $\couplingstrength$. Second, the first order energy corrections are proportional to the eigenvalues of $\projectordegenspace \interaction \projectordegenspace$. Again, we assume here that these corrections lift the degeneracy, i.e., the corrections $\energyglobal^{(1)}_m \neq \energyglobal^{(1)}_n$ differ for $m \neq n$. As a result, a new splitting 
\begin{subequations}
\begin{align}
    \hamiltonianglobal 
    &= \left( \hamiltonianglobalfree + \couplingstrength \projectordegenspace \interaction \projectordegenspace \right) + \couplingstrength \left( \interaction - \projectordegenspace \interaction \projectordegenspace \right) \\
    &= \hamiltonianglobalfree' + \couplingstrength \interaction'
\end{align}
\end{subequations}
of the Hamiltonian can be used. Now, the new Hamiltonian $\hamiltonianglobalfree'$ does not contain degeneracies in the subspace of $\projectordegenspace$. As remarked above, we choose one of the eigenstates of $\projectordegenspace \interaction \projectordegenspace$ as the zeroth order state, such that
\begin{equation}
    \hamiltonianglobalfree' \ketglobal{\stateglobal^{(0)}} 
    = \energyglobal'^{(0)} \ketglobal{\stateglobal^{(0)}} \,,
\end{equation}
where the zeroth order energy $\energyglobal'^{(0)}$ contains already a first order correction. Applying the first order scheme to this new splitting yields
\begin{equation}
    \couplingstrength \ketglobal{\stateglobal^{(1)}} 
    = \couplingstrength \projectordegenspacecomp \Bigl( \energyglobal^{(0)} - \hamiltonianglobalfree \Bigr)^{-1} \projectordegenspacecomp \interaction \ketglobal{\stateglobal^{(0)}} + \mathcal{O}(\couplingstrength^2) \,,
\end{equation}
matching the previous expression~\eqref{eq:global_state_correction}. However, the operator $\projectordegenspacecomp ( \energyglobal^{(0)} - \hamiltonianglobalfree )^{-1} \projectordegenspacecomp$ does not include states which lie in the degenerate subspace of the original problem. One naturally asks how the other eigenstates contribute to the first order state correction. 
Indeed, they have a non-vanishing influence and show up in the next set of perturbation terms, because the inverse energy difference $(\energyglobal'^{(0)} - \hamiltonianglobalfree')^{-1}$ scales with $1/\couplingstrength$ for these states. Without explicitly deriving this commonly known term~\cite{Picasso2022}, 
we simply state
\begin{align}
    \couplingstrength^2 \ketglobal{\stateglobal^{(2)}} 
    &= \couplingstrength \projectordegenspacewozero \Bigl[ \melglobal{\stateglobal^{(0)}}{\interaction}{\stateglobal^{(0)}} - \projectordegenspace \interaction \projectordegenspace \Bigr]^{-1} \projectordegenspacewozero \interaction \ketglobal{\stateglobal^{(1)}} \nonumber\\
    &\quad + \mathcal{O}(\couplingstrength^2)
    \label{eq:degeneracy_second_order_scheme_state_correction}
\end{align}
where $\energyglobal'^{(1)} = \melglobal{\stateglobal^{(0)}}{\interaction'}{\stateglobal^{(0)}} = 0$. In the last line, we introduce the projector $\projectordegenspacewozero = \projectordegenspace - \projectorglobal^{(0)}$ onto the degenerate subspace of $\energyglobal^{(0)}$ excluding the one-dimensional space corresponding to the zeroth order global state $\ketglobal{\stateglobal^{(0)}}$. Although this correction has an influence on the initial system state, it does not change the first order correction in the TDPT scheme, because the original degeneracy manifests as
\begin{eqnarray}
   \Bigl( \energyglobal^{(0)} - \hamiltonianglobalfree \Bigr) \projectordegenspacewozero 
   = 0
\end{eqnarray}
in Eq.~\eqref{eq:energy_relation_first_order_correction_for_time_integral}. 
Hence, Eq.~\eqref{eq:final_result_tdpt} remains unchanged in this situation. 
If the degeneracy is not lifted at first order, then higher orders must be used until the degeneracy finally disappears (unless the perturbation $\interaction$ exhibits the same symmetry as the unperturbed Hamiltonian $\hamiltonianglobalfree$).

%%%%%%%%%%%%%%%%%%%%%%%%%%%%%%%%%%%%%%%%%%%%%%%%%%%%%%%%%%%%%%%%%%%%%%%%%%%%%%%%

\section{Examples}

\begin{figure}
    \centering
    \pgfmathsetmacro{\anglespin}{-30}
    \pgfmathsetmacro{\anglespinj}{45}
    \pgfmathsetmacro{\lengthspin}{0.56}
    \pgfmathsetmacro{\lengthspinj}{1.0}
    \pgfmathsetmacro{\facright}{0.74}
    \pgfmathsetmacro{\facrightdown}{1}
    \pgfmathsetmacro{\facrightup}{2.0}
    \pgfmathsetmacro{\cosanglespin}{cos(\anglespin)}	
    \pgfmathsetmacro{\sinanglespin}{sin(\anglespin)}
    \pgfmathsetmacro{\cosanglespinj}{cos(\anglespinj)}	
    \pgfmathsetmacro{\sinanglespinj}{sin(\anglespinj)}
    \pgfmathsetmacro{\xleftspin}{-\lengthspin * \sinanglespin}
    \pgfmathsetmacro{\yleftspin}{\lengthspin * \cosanglespin}
    \pgfmathsetmacro{\xrightspin}{\facright * \lengthspin * \sinanglespin}
    \pgfmathsetmacro{\yrightspin}{-\facright * \lengthspin * \cosanglespin}
    \pgfmathsetmacro{\xleftspinj}{-\facrightup * \lengthspin * \sinanglespinj}
    \pgfmathsetmacro{\yleftspinj}{\facrightup * \lengthspin * \cosanglespinj}
    \pgfmathsetmacro{\xrightspinj}{\facrightdown * \lengthspinj * \sinanglespinj}
    \pgfmathsetmacro{\yrightspinj}{-\facrightdown * \lengthspinj * \cosanglespinj}
    \pgfmathsetmacro{\xspina}{-1.3}
    \pgfmathsetmacro{\xspinb}{-\xspina}
    \pgfmathsetmacro{\midline}{1.5}
    \pgfmathsetmacro{\yoffsetlabel}{0.9}
    \pgfmathsetmacro{\radiusspin}{0.22}
    \pgfmathsetmacro{\radiusspinj}{0.42}
    \begin{tikzpicture}[scale=1.0]
    	% 
    	% ENTANGLEMENT
    	%
    	\draw[draw=gray!80!black, line width=1.3pt, solid, snakeit] ({\xspina}, {\midline}) to[bend right=30] ({\xspinb}, {\midline});
    	%
    	% SPINS
    	%
    	% system
    	\draw[{Latex[width=3mm, length=2.6mm]}-, draw=black!100, line width=2.4pt, solid] ({\xspina + \xleftspin},{\midline + \yleftspin}) -- ({\xspina + \xrightspin},{\midline + \yrightspin});
    	\shade[ball color=red!80] ({\xspina},{\midline}) circle ({\radiusspin});
    	%
    % 	
    	%
    	% clock
    	\draw[{Latex[width=3mm, length=2.6mm]}-, draw=black!100, line width=2.9pt, solid] ({\xspinb + \xleftspinj},{\midline + \yleftspinj}) -- ({\xspinb + \xrightspinj},{\midline + \yrightspinj});
    	\shade[ball color=blue!80] ({\xspinb},{\midline}) circle ({\radiusspinj});
    	%
    	% LABELS
    	\node (env) at (\xspina, {\midline-\yoffsetlabel}) [black] {\textbf{S}};
    	\node (env) at (\xspinb, {\midline-\yoffsetlabel}) [black] {\textbf{C}};
    \end{tikzpicture}
    \caption{For a demonstration of the relational approach in perturbation theory, a spin-1/2 (red) is coupled to an environment, taken as a large spin (blue).}
    \label{fig:sketch_example_setting}
\end{figure}
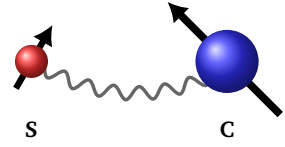

For a quantitative analysis of our derived results, we consider a physical setting of two coupled spins (see Fig.~\ref{fig:sketch_example_setting}). 
In particular, the system is taken as a spin-1/2, while the environment comprises a large spin with fixed spin $J$. Both subsystems are described by the sum 
\begin{equation}
    \hamiltonianglobalfree 
    = \energysystem \sigmaz \otimes \identclock + \energyclock \identsystem \otimes \angmomz 
    \label{eq:example_unperturbed_global_Hamiltonian}
\end{equation}
of their native Hamiltonians $\hamiltoniansystem_0 = \energysystem \sigmaz$ and $\hamiltonianclock = \energyclock \angmomz$ at the initial stage. Pauli matrices $\sigmai$ and angular momentum operators $\angmomi$ for $i\in\{x,y,z\}$ take their usual forms. The uncoupled Hamiltonians provide a natural basis for each part, namely $\{\ketsystem{\uparrow}, \ketsystem{\downarrow} \}$ for the system and $\{ \ketclock{m} \}$ with $m \in [ -J, \dots , J]$ for the clock. Each subsystem is characterized by their individual energy scales $\energysystem$ and $\energyclock$ for system and clock, respectively. In our example, the perturbation in $\hamiltonianglobal = \hamiltonianglobalfree + \couplingstrength \interaction$ takes the explicit form 
\begin{equation}
    \interaction 
    = \sigmax \otimes \hat{F} \,,
\end{equation}
where the clock operator $\hat{F}$ can be arbitrary. The purpose of our example is a comprehensible and transparent demonstration of the derived result. To this end, we choose
\begin{equation}
    \melclock{m}{\hat{F}}{n} 
    = 1 \qquad \forall m,n
\end{equation}
as it provides a straightforward way to analytical expressions without complicated terms obscuring the working of our method. 

Considering only few-level systems admits straightforward analytical solutions, but limits the complexity of the obtainable system dynamics as a tradeoff. In particular, the inherent periodicity $T = 2\pi/\energyclock$ of arbitrary clock states imprints the same cyclicity on the system, i.e., $\ketsystem{\statesystem(T)} = \ketsystem{\statesystem(0)}$. We examine two scenarios, a non-degenerate and a degenerate one, in order to illustrate the correctness of our results independent of degeneracy. 

%%%%%%%%%%%%%%%%%%%%%%%%%%%%%%%%%%%%

\subsection{No degeneracy}

\begin{figure}
    \centering
    \includegraphics[width=0.5\textwidth]{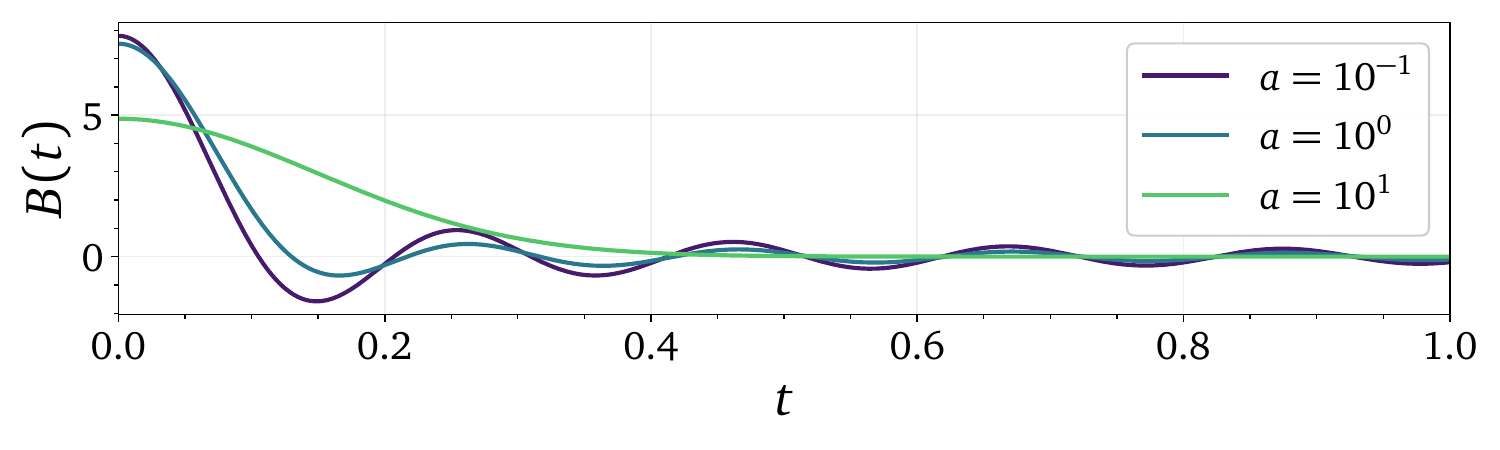}
    \caption{The real-valued function $B(\prm) = b_0 + 2\sum_{m=1}^{J} b_m \cos(m\energyclock \prm)$ with $b_m \propto \exp[-a(m/J)^2]$ is shown for $J=30$ and $\energyclock = 1$. Different colors indicate three widths $a \in \{ 0.1, 1, 10\}$.}
    \label{fig:example_B_t}
\end{figure}

By restraining the system energy scale $\energysystem$ to non-integer multiples of $\energyclock$, we ensure the non-degeneracy of the unperturbed spectrum. To be specific, we select the unperturbed global state $\ketglobal{\stateglobal^{(0)}} = \ketglobal{\uparrow, m=0}$ with energy $\energyglobal^{(0)} = \energysystem$ and obtain
\begin{equation}
    \interaction \ketglobal{\stateglobal^{(0)}} 
    = \ketsystem{\downarrow} \otimes \sum_{m=-J}^{J} \ketclock{m} \,.
\end{equation}
As a result, the first order perturbations read $\energyglobal^{(1)} = 0$ and
\begin{equation}
    \ketglobal{\stateglobal^{(1)}}
    = \sum_{m=-J}^{J} \frac{\ketglobal{\downarrow,m}}{2\energysystem-m\energyclock} \,,
\end{equation}
changing only the global state. We express the clock state
$\ketclock{\stateclock(\prm)} = e^{i\prm\energyglobal} \sum_{m=-J}^{J} b_m e^{-i\prm \energyclock m} \ketclock{m}$, in the eigenbasis of the clock Hamiltonian $\hamiltonianclock$ and, in particular, find the initial state
\begin{equation}
    \ketsystem{\statesystem(0)} 
    \approx \frac{1}{\abs{b_0}} \left[ b_0^* \ketsystem{\uparrow} + \couplingstrength \ketsystem{\downarrow} \sum_{m=-J}^{J} \frac{b_m^*}{2\energysystem - m\energyclock} \right] \,.
    \label{eq:example_nondegenerate_initial_system_state}
\end{equation}
In a direct manner, one further evaluates $\melclockglobal{\stateclock(\prm)}{\interaction}{\stateglobal^{(0)}} = e^{-i\energyglobal\prm} \ketsystem{\downarrow} \cdot B(\prm)$ with $B(\prm) = \sum_{m=-J}^{J} b_m^* e^{i\prm m}$ and, hence, obtains the effective system potential
\begin{subequations}
\begin{align}
    \potentialsystem^{(0)}(\prm) 
    &= \frac{1}{\abs{b_0}^2} \Biggl[ b_0 B(\prm) \dyadsystemop{\downarrow}{\uparrow} + b_0^* B^*(\prm) \dyadsystemop{\uparrow}{\downarrow} \Biggr] \\
    &= \frac{ \Re \bigl[ b_0 B(\prm) \bigr] \, \sigmax + \Im \bigl[ b_0 B(\prm) \bigr] \, \sigmay  }{\abs{b_0}^2} \,.
    \label{eq:example_nondegenerate_effective_potential}
\end{align}
\end{subequations}
As a special case, we assume $b_{-m} = b_{m}$ and $b_{m} \geq 0$ real for all $m$, which renders $B(\prm)$ real and the effective potential simplifies to
\begin{equation}
    \potentialsystem^{(0)}(\prm)
    = \frac{B(\prm)}{b_0} \sigmax \,.
\end{equation}
The prefactor $B(\prm)$ dictates the time-dependence of the effective potential and implies that a broad clock energy distribution will result in a finite time period for which the potential significantly differs from zero. This behavior is illustrated in Fig.~\ref{fig:example_B_t} for Gaussian clock energy distributions $b_m \propto \exp[-a(m/J)^2]$ with $a>0$. 
Equations~\eqref{eq:example_nondegenerate_initial_system_state} and~\eqref{eq:example_nondegenerate_effective_potential} clearly demonstrate the necessity for a non-vanishing clock population of the state $\ketclock{m=0}$ in order to obtain non-trivial system dynamics. Such a characteristic is a general feature of the relational approach to time, because, roughly speaking, the clock state must sample as much as possible of the entanglement contained in the global state~\cite{Gemsheim2023phd} for intricate system evolutions~\cite{Gemsheim2023a}.

\begin{figure}
    \centering
    \includegraphics[width=0.5\textwidth]{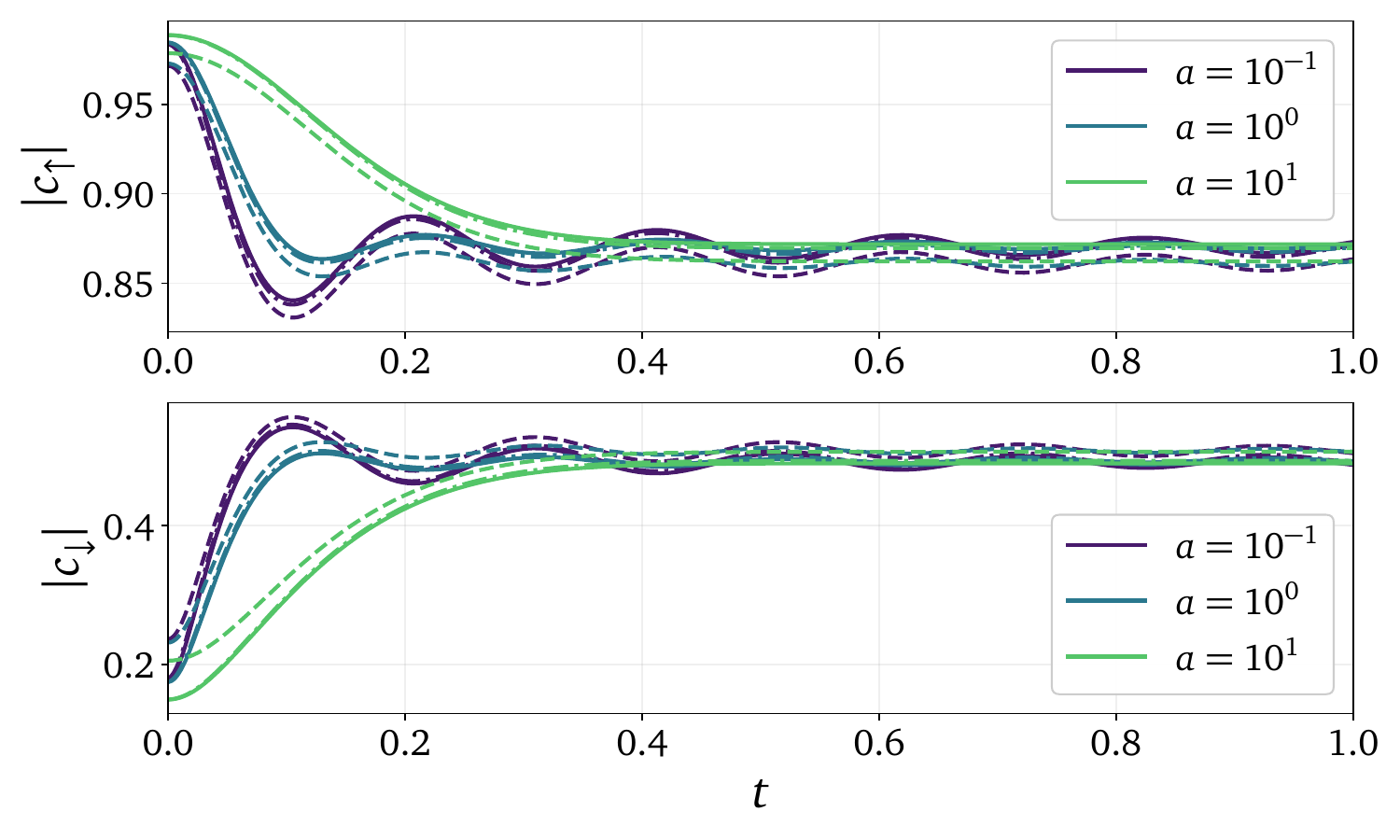}
    \caption{Both, upper and lower level population dynamics of the primary system are shown in the upper and lower panel, respectively, for $J=30$, $\couplingstrength=0.17$, $\energysystem=0.32$ and $\energyclock = 1$. Such a choice of parameters corresponds to the non-degenerate scenario. 
    Relational system states derived from the full numerical (solid) and TIPT solution (dashed) for the global state is compared to the TDPT result (dash-dotted). 
    }
    \label{fig:example_nondegenerate_system_dynamics}
\end{figure}

A comparison of the relational dynamics for the exact and approximate global state, as well as the first order TDPT solution, is shown in Fig.~\ref{fig:example_nondegenerate_system_dynamics} for a specific set of parameters with the aforementioned Gaussian clock distribution. 

%%%%%%%%%%%%%%%%%%%%%%%%%%%%%%%%%%%%

\subsection{Degeneracy}

\begin{figure}
    \centering
    \includegraphics[width=0.5\textwidth]{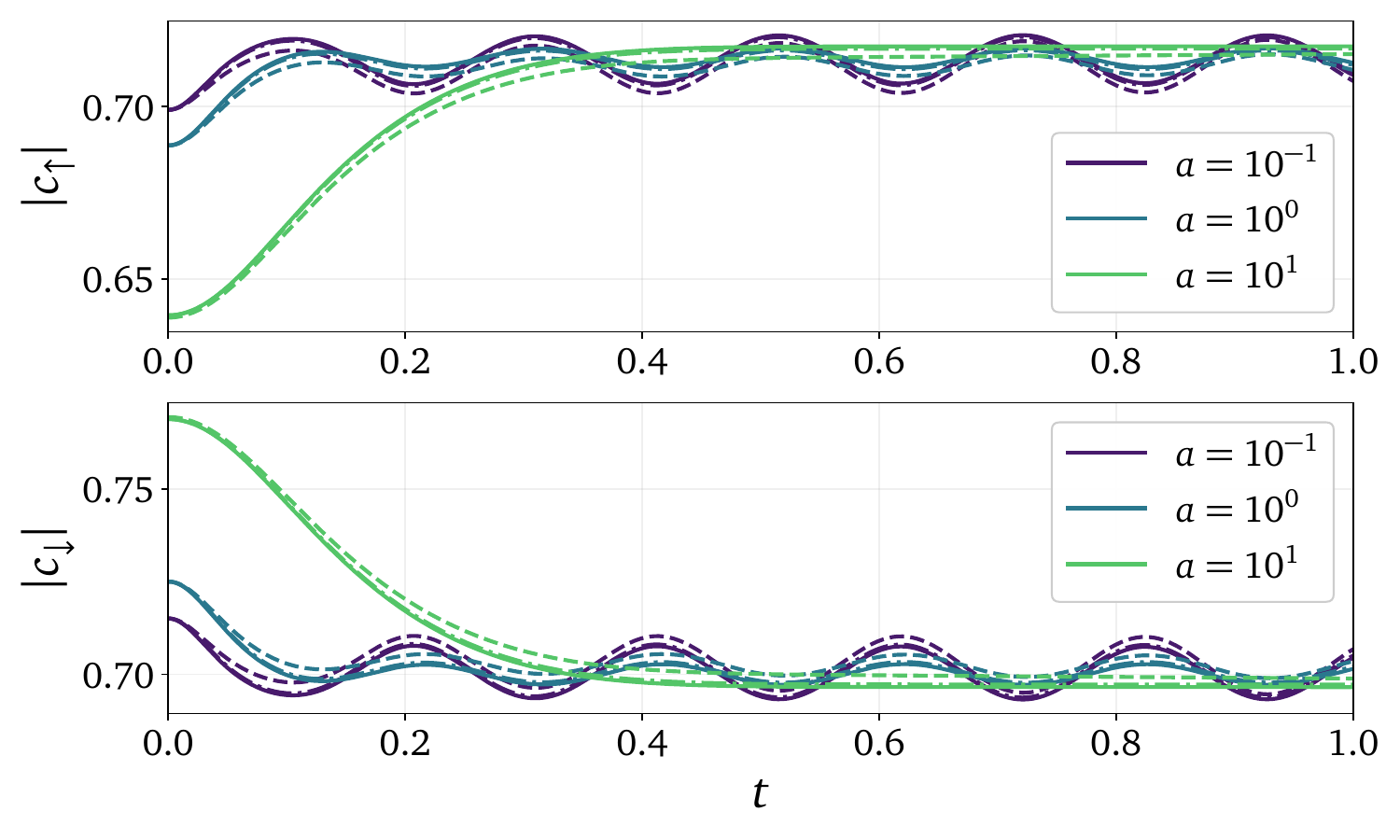}
    \caption{The dynamics of the system is shown for the degenerate case with $J=30$, $\couplingstrength=0.29$ and $\energysystem=\energyclock=1$. Plot elements are the same as in Fig.~\ref{fig:example_degenerate_system_dynamics}.}
    \label{fig:example_degenerate_system_dynamics}
\end{figure}

The aforementioned physical settings can host a degenerate spectrum as well. Almost all energy levels become doubly degenerate by setting the system energy scale to $\energysystem = \energyclock$. As a specific choice, we consider the unperturbed energy level $\energyglobal^{(0)} = 0$ with the eigenvectors
\begin{equation}
    \Bigl\{ \ketglobal{\uparrow, m=-1}, \ketglobal{\downarrow, m=1} \Bigr\} \,.
\end{equation}
Since this level is degenerate, we have to check if the degeneracy is lifted at first order~\cite{Picasso2022} in the subspace projected by 
\begin{equation}
    \projectordegenspace 
    = \dyadglobal{\uparrow,-1}{\uparrow,-1} + \dyadglobal{\downarrow,1}{\downarrow,1} \,.
\end{equation}
To this end, we determine the eigenvalues and -vectors of the subspace block  
\begin{equation}
    \projectordegenspace \interaction \projectordegenspace 
    = \dyadglobal{\uparrow,-1}{\downarrow,1} + \dyadglobal{\downarrow,1}{\uparrow,-1} \,.
    \label{eq:degenerate_subspace_projected_interaction}
\end{equation}
of the interaction to be $\lambda_{\pm} = \pm 1$ and
\begin{equation}
    \ketglobal{\beta_{\pm}} 
    = \frac{1}{\sqrt{2}} \Bigl( \ketglobal{\uparrow,-1} \pm \ketglobal{\downarrow,1} \Bigr) \,,
\end{equation}
respectively. The zeroth order for the global state is taken as the plus sign solution $\ketglobal{\stateglobal^{(0)}} = \ketglobal{\beta_{+}}$ with the eigenenergy $\energyglobal'^{(0)} = \couplingstrength $. Surprisingly, the operation of the global interaction disentangles the original $\ketglobal{\stateglobal^{(0)}}$, i.e., $\interaction \ketglobal{\stateglobal^{(0)}} = \frac{1}{\sqrt{2}} \Bigl( \ketsystem{\downarrow} + \ketsystem{\uparrow}  \Bigr) \otimes \sum_{m=-J}^{J} \ketclock{m}$, and yields the first order term
\begin{equation}
    \ketglobal{\stateglobal^{(1)}} 
    = \frac{-1}{\sqrt{2}\energyclock} \left[ \sum_{m\neq1} \frac{ \ketglobal{\downarrow,m} }{m-1} + \sum_{m\neq-1} \frac{ \ketglobal{\uparrow,m} }{m+1} \right]
\end{equation}
in the perturbation scheme. A straightforward calculation with $\projectordegenspacewozero = \dyadglobal{\beta_{-}}{\beta_{-}}$ yields
\begin{subequations}
\begin{align}
    \couplingstrength \ketglobal{\stateglobal^{(2)}} 
    &= \frac{1}{2} \ketglobal{\beta_{-}} \melglobal{\beta_{-}}{ \interaction }{\stateglobal^{(1)}} \\
    &= - \frac{1}{4\energyclock} \ketglobal{\beta_{-}} \left[ \sum_{m\neq1} \frac{ 1 }{m-1} - \sum_{m\neq-1} \frac{ 1 }{m+1} \right] \\
    &= -\frac{1}{2\energyclock} \ketglobal{\beta_{-}} \sum_{m=-J, m\neq1}^{J} \frac{1}{m-1} \\
    &= \frac{G}{\energyclock} \ketglobal{\beta_{-}} 
\end{align}
\end{subequations}
with $G = [ 1/J + 1/(J+1) ]/2 \rightarrow 1/J$ for $J \gg 1$. Putting everything together, the global state reads
\begin{align}
    \ketglobal{\stateglobal} 
    &\approx \frac{\ketsystem{\uparrow} }{\sqrt{2}} \otimes \Biggl[ \left( 1 + \frac{\couplingstrength G}{\energyclock} \right) \ketclock{-1} - \frac{\couplingstrength}{\energyclock} \sum_{m\neq-1} \frac{ \ketclock{m} }{m+1} \Biggr] \nonumber\\
    &\quad + \frac{\ketsystem{\downarrow} }{\sqrt{2}} \otimes \Biggl[ \left( 1 - \frac{\couplingstrength G}{\energyclock} \right) \ketclock{1} - \frac{\couplingstrength}{\energyclock} \sum_{m\neq1} \frac{ \ketclock{m}}{m-1} \Biggr]
\end{align}
in first order of $\couplingstrength$. The details of the calculation for the zeroth order effective potential have been moved to App.~\ref{app:calculations_effective_potential_degenerate} for a more comprehensible presentation. We simply give the result as 
\begin{align}
    \potentialsystem^{(0)}(\prm)
    &= \frac{1}{\abs{b_{-1}}^2 + \abs{b_{1}}^2} \Biggl\{ \Re \Bigl[ B(\prm) \Bigl( b_{1} e^{-i\prm\energyclock} - b_{-1} e^{i\prm\energyclock} \Bigr) \Bigr] \sigmaz \nonumber\\
    &\qquad\quad + \Re \Bigl[ e^{i\prm\energyclock} \Bigl(  b_{-1} B(\prm) + b_{1}^*B^*(\prm) \Bigr) \Bigr] \sigmax \nonumber\\
    &\qquad\quad + \Im \Bigl[ e^{i\prm\energyclock} \Bigl(  b_{-1} B(\prm) + b_{1}^*B^*(\prm) \Bigr) \Bigr] \sigmay \Biggr\} \,.
    \label{eq:effective_potential_degenerate_general}
\end{align}
Using again the simplifying assumptions about symmetric, positive coefficients $b_m$, the last expression reduces to
\begin{equation}
    \potentialsystem^{(0)}(\prm)
    = \frac{B(\prm)}{b_{1}} \Bigl[ \cos(\prm\energyclock) \sigmax + \sin(\prm\energyclock) \sigmay \Bigr] \,.
\end{equation}
Again, we visualize the dynamics of the system (in the Schr{\"o}dinger picture) for an exemplary set of parameters in Fig.~\ref{fig:example_degenerate_system_dynamics}. 

In both cases, degenerate and non-degenerate, the temporal evolutions of the system for relational TIPT and TDPT adequately match each other. Hence, our numerical demonstrations in this section complement our analytical results and further reinforce their correctness. Although simple, these examples provide sufficient insight into the workings of the discovered connection between perturbation theory on a static and dynamical level.

%%%%%%%%%%%%%%%%%%%%%%%%%%%%%%%%%%%%%%%%%%%%%%%%%%%%%%%%%%%%%%%%%%%

\section{Conclusion and Outlook}

In this work, we derived the relational state of a system from the perturbed energy eigenstate of an enlarged system, additionally comprising the environment it is coupled to. Based on these findings, we have successfully established that the elements of time-dependent perturbation theory can emerge from its static counterpart, the time-independent perturbation theory. In other words, the deep link between the static and dynamical realm within the relational framework becomes also evident within perturbation theory, one of the most important tools in physics. 

Our result in first order of the coupling show that exact TDPT expressions can be obtained without the need to solve time integrals. If the effective potential~\eqref{eq:effective_potential_zeroth_order} matches a desired one and an efficient calculation of the inverse in Eq.~\eqref{eq:global_state_correction} exists, then it may prove useful as an alternative way to compute dynamical corrections. For weak couplings, even other methods for resolving parts of the global spectrum, such as in Refs.~\cite{Polizzi2009, Sierant2020}, can be used for the determination of global eigenstates. The perturbed system dynamics, yielded from these solutions, are still correct to first order. 

At the current stage, our finding holds conceptual value as a non-trivial connection between distinct frameworks for perturbations. Further calculations of more realistic physical scenarios will prove useful for practical applications. Given that relational time carries over to the purely classical realm~\cite{Vedral2023, Gemsheim2023phd}, we also anticipate our conclusion to hold for classical states on phase space. Lastly, providing a general scheme up to arbitrary orders serves as a noteworthy research direction for future work.

%%%%%%%%%%%%%%%%%%%%%%%%%%%%%%%%%%%%%%%%%%%%%%%%%%%%

\subsection{Acknowledgements}

Feedback from Ulf Saalmann and Jan-Michael Rost regarding the manuscript is greatly appreciated. Furthermore, we thank Benjamin Rabe for discussions.

%%%%%%%%%%%%%%%%%%%%%%%%%%%%%%%%%%%%%%%%%%%%%%%%%%%%

\appendix

\section{Effective potential for degenerate case}
\label{app:calculations_effective_potential_degenerate}

Some essential elements for the calculation of the effective system potential $\potentialsystem^{(0)}$ in zeroth order are
\begin{subequations}
\begin{align}
    \braketclockglobal{\stateclock(\prm)}{\stateglobal^{(0)}} 
    &= \frac{e^{-i\prm\energyglobal}}{\sqrt{2}} \Bigl(  b_{-1}^* e^{-i\prm\energyclock} \ketsystem{\uparrow} + b_{1}^* e^{i\prm\energyclock} \ketsystem{\downarrow}  \Bigr) \,,\\
    \melclockglobal{\stateclock(\prm)}{\interaction}{\stateglobal^{(0)}} 
    &= \frac{B(\prm) \, e^{-i\prm\energyglobal} }{\sqrt{2}} \Biggl( \ketsystem{\downarrow} + \ketsystem{\uparrow} \Biggr) 
\end{align}
\end{subequations}
and
\begin{equation}
    \melglobal{\stateglobal^{(0)}}{\projectorclockinit}{\stateglobal^{(0)}} 
    = \frac{\abs{b_{-1}}^2 + \abs{b_{1}}^2}{2} \,.
\end{equation}
Based on these basic expression, one immediately finds the scalar
\begin{equation}
    \melglobal{\stateglobal^{(0)}}{\interaction\projectorclock(\prm)}{\stateglobal^{(0)}}
     = \frac{B^*(\prm) }{2}  \Bigl( b_1^* e^{i\prm\energyclock} + b_{-1}^*  e^{-i\prm\energyclock} \Bigr) 
    \label{eq:normalization_zeroth_order_degenerate}
\end{equation}
and the non-hermitian system operator
\begin{align}
    &\melclock{\stateclock(\prm)}{ \interaction \projectorglobal^{(0)} }{\stateclock(\prm)} \nonumber\\
    &\qquad = \frac{B(\prm) }{2} \Bigl( \ketsystem{\downarrow} + \ketsystem{\uparrow}  \Bigr) \Bigl( b_{-1} e^{i\prm\energyclock} \brasystem{\uparrow} +  b_1 e^{-i\prm\energyclock} \brasystem{\downarrow}  \Bigr) \,.
\end{align}
The last line can be symmetrized to 
\begin{subequations}
\begin{align}
    &\hspace{-4mm} \melclock{\stateclock(\prm)}{\Bigl\{ \interaction, \projectorglobal^{(0)} \Bigr\} }{\stateclock(\prm)} \nonumber\\
    & = \frac{B(\prm) }{2} \Bigl( \ketsystem{\downarrow} + \ketsystem{\uparrow}  \Bigr) \Bigl( b_{-1} e^{i\prm\energyclock} \brasystem{\uparrow} + b_{1}  e^{-i\prm\energyclock} \brasystem{\downarrow}  \Bigr) \nonumber\\
    &\quad + \frac{B^*(\prm) }{2} \Bigl( b_{-1}^* e^{-i\prm\energyclock} \ketsystem{\uparrow} + b_{1}^* e^{i\prm\energyclock} \ketsystem{\downarrow} \Bigr) \Bigl( \brasystem{\uparrow} + \brasystem{\downarrow} \Bigr) \\
    &= \frac{1}{2} \Biggl\{ \dyadsystemop{\downarrow}{\downarrow} \Bigl[ b_{1} B(\prm) e^{-i\prm\energyclock} + b_{1}^*B^*(\prm) e^{i\prm\energyclock}  \Bigr] \nonumber\\
    &\quad + \dyadsystemop{\uparrow}{\uparrow} \Bigl[ b_{-1} B(\prm) e^{i\prm\energyclock} + b_{-1}^*B^*(\prm) e^{-i\prm\energyclock}  \Bigr] \nonumber\\
    &\quad + \dyadsystemop{\uparrow}{\downarrow} \, e^{-i\prm\energyclock} \Bigl[ b_{1} B(\prm) + b_{-1}^*B^*(\prm) \Bigr] \nonumber\\
    &\quad + \dyadsystemop{\downarrow}{\uparrow} \, e^{i\prm\energyclock} \Bigl[ b_{-1} B(\prm) + b_{1}^*B^*(\prm) \Bigr] \Biggr\} \\
    & = \dyadsystemop{\downarrow}{\downarrow} \Re \Bigl[ b_{1} B(\prm) e^{-i\prm\energyclock} \Bigr]  + \dyadsystemop{\uparrow}{\uparrow} \Re \Bigl[ b_{-1} B(\prm) e^{i\prm\energyclock} \Bigr] \nonumber\\
    &\quad + \dyadsystemop{\uparrow}{\downarrow} \, e^{-i\prm\energyclock} \frac{b_{1} B(\prm) + b_{-1}^*B^*(\prm)}{2} \nonumber\\
    &\quad + \dyadsystemop{\downarrow}{\uparrow} \, e^{i\prm\energyclock} \frac{ b_{-1} B(\prm) + b_{1}^*B^*(\prm) }{2} \\
    & = \frac{1}{2} \Biggl\{ \Re \left[ B(\prm) \Bigl( b_{1} e^{-i\prm\energyclock} + b_{-1} e^{i\prm\energyclock} \Bigr) \right] \ident \nonumber\\
    &\quad + \Re \Bigl[ B(\prm) \Bigl( b_{1} e^{-i\prm\energyclock} - b_{-1} e^{i\prm\energyclock} \Bigr) \Bigr] \sigmaz \nonumber\\
    &\quad + \Re \Bigl[ e^{i\prm\energyclock} \Bigl(  b_{-1} B(\prm) + b_{1}^*B^*(\prm) \Bigr) \Bigr] \sigmax \nonumber\\
    &\quad + \Im \Bigl[ e^{i\prm\energyclock} \Bigl(  b_{-1} B(\prm) + b_{1}^*B^*(\prm) \Bigr) \Bigr] \sigmay \Biggr\}
\end{align}
\end{subequations}
and yields Eq.~\eqref{eq:effective_potential_degenerate_general} of the main text after the inclusion of the normalization factor~\eqref{eq:normalization_zeroth_order_degenerate}.

%%%%%%%%%%%%%%%%%%%%%%%%%%%%%%%%%%%%%%%%%%%%%%%%%%%%

\bibliography{bib_link_perturbation}

\begin{thebibliography}{35}%
\makeatletter
\providecommand \@ifxundefined [1]{%
 \@ifx{#1\undefined}
}%
\providecommand \@ifnum [1]{%
 \ifnum #1\expandafter \@firstoftwo
 \else \expandafter \@secondoftwo
 \fi
}%
\providecommand \@ifx [1]{%
 \ifx #1\expandafter \@firstoftwo
 \else \expandafter \@secondoftwo
 \fi
}%
\providecommand \natexlab [1]{#1}%
\providecommand \enquote  [1]{``#1''}%
\providecommand \bibnamefont  [1]{#1}%
\providecommand \bibfnamefont [1]{#1}%
\providecommand \citenamefont [1]{#1}%
\providecommand \href@noop [0]{\@secondoftwo}%
\providecommand \href [0]{\begingroup \@sanitize@url \@href}%
\providecommand \@href[1]{\@@startlink{#1}\@@href}%
\providecommand \@@href[1]{\endgroup#1\@@endlink}%
\providecommand \@sanitize@url [0]{\catcode `\\12\catcode `\$12\catcode
  `\&12\catcode `\#12\catcode `\^12\catcode `\_12\catcode `\%12\relax}%
\providecommand \@@startlink[1]{}%
\providecommand \@@endlink[0]{}%
\providecommand \url  [0]{\begingroup\@sanitize@url \@url }%
\providecommand \@url [1]{\endgroup\@href {#1}{\urlprefix }}%
\providecommand \urlprefix  [0]{URL }%
\providecommand \Eprint [0]{\href }%
\providecommand \doibase [0]{https://doi.org/}%
\providecommand \selectlanguage [0]{\@gobble}%
\providecommand \bibinfo  [0]{\@secondoftwo}%
\providecommand \bibfield  [0]{\@secondoftwo}%
\providecommand \translation [1]{[#1]}%
\providecommand \BibitemOpen [0]{}%
\providecommand \bibitemStop [0]{}%
\providecommand \bibitemNoStop [0]{.\EOS\space}%
\providecommand \EOS [0]{\spacefactor3000\relax}%
\providecommand \BibitemShut  [1]{\csname bibitem#1\endcsname}%
\let\auto@bib@innerbib\@empty
%</preamble>
\bibitem [{\citenamefont {Wigner}(1960)}]{Wigner1960}%
  \BibitemOpen
  \bibfield  {author} {\bibinfo {author} {\bibfnamefont {E.~P.}\ \bibnamefont
  {Wigner}},\ }\bibfield  {title} {\bibinfo {title} {{The unreasonable
  effectiveness of mathematics in the natural sciences. Richard courant lecture
  in mathematical sciences delivered at New York University, May 11, 1959}},\
  }\href {https://doi.org/10.1002/cpa.3160130102} {\bibfield  {journal}
  {\bibinfo  {journal} {Communications on Pure and Applied Mathematics}\
  }\textbf {\bibinfo {volume} {13}},\ \bibinfo {pages} {1} (\bibinfo {year}
  {1960})}\BibitemShut {NoStop}%
\bibitem [{\citenamefont
  {Schr\"{o}dinger}(1926{\natexlab{a}})}]{Schroedinger1926_3}%
  \BibitemOpen
  \bibfield  {author} {\bibinfo {author} {\bibfnamefont {E.}~\bibnamefont
  {Schr\"{o}dinger}},\ }\bibfield  {title} {\bibinfo {title} {{Quantisierung
  als Eigenwertproblem}},\ }\href {https://doi.org/10.1002/andp.19263851302}
  {\bibfield  {journal} {\bibinfo  {journal} {Annalen der Physik}\ }\textbf
  {\bibinfo {volume} {385}},\ \bibinfo {pages} {437} (\bibinfo {year}
  {1926}{\natexlab{a}})}\BibitemShut {NoStop}%
\bibitem [{\citenamefont
  {Schr\"{o}dinger}(1926{\natexlab{b}})}]{Schroedinger1926_4}%
  \BibitemOpen
  \bibfield  {author} {\bibinfo {author} {\bibfnamefont {E.}~\bibnamefont
  {Schr\"{o}dinger}},\ }\bibfield  {title} {\bibinfo {title} {{Quantisierung
  als Eigenwertproblem}},\ }\href {https://doi.org/10.1002/andp.19263861802}
  {\bibfield  {journal} {\bibinfo  {journal} {Annalen der Physik}\ }\textbf
  {\bibinfo {volume} {386}},\ \bibinfo {pages} {109} (\bibinfo {year}
  {1926}{\natexlab{b}})}\BibitemShut {NoStop}%
\bibitem [{\citenamefont {Sambe}(1973)}]{Sambe1973}%
  \BibitemOpen
  \bibfield  {author} {\bibinfo {author} {\bibfnamefont {H.}~\bibnamefont
  {Sambe}},\ }\bibfield  {title} {\bibinfo {title} {{Steady States and
  Quasienergies of a Quantum-Mechanical System in an Oscillating Field}},\
  }\href {https://doi.org/10.1103/PhysRevA.7.2203} {\bibfield  {journal}
  {\bibinfo  {journal} {Phys. Rev. A}\ }\textbf {\bibinfo {volume} {7}},\
  \bibinfo {pages} {2203} (\bibinfo {year} {1973})}\BibitemShut {NoStop}%
\bibitem [{\citenamefont {Grossmann}(2018)}]{Grossmann2018}%
  \BibitemOpen
  \bibfield  {author} {\bibinfo {author} {\bibfnamefont {F.}~\bibnamefont
  {Grossmann}},\ }\href@noop {} {\emph {\bibinfo {title} {{{Theoretical
  femtosecond physics}}}}},\ \bibinfo {edition} {3rd}\ ed.,\ Graduate texts in
  physics\ (\bibinfo  {publisher} {Springer International Publishing},\
  \bibinfo {address} {Cham, Switzerland},\ \bibinfo {year} {2018})\BibitemShut
  {NoStop}%
\bibitem [{\citenamefont {del Campo}\ \emph {et~al.}(2017)\citenamefont {del
  Campo}, \citenamefont {Molina-Vilaplana},\ and\ \citenamefont
  {Sonner}}]{delCampo2017}%
  \BibitemOpen
  \bibfield  {author} {\bibinfo {author} {\bibfnamefont {A.}~\bibnamefont {del
  Campo}}, \bibinfo {author} {\bibfnamefont {J.}~\bibnamefont
  {Molina-Vilaplana}},\ and\ \bibinfo {author} {\bibfnamefont {J.}~\bibnamefont
  {Sonner}},\ }\bibfield  {title} {\bibinfo {title} {{Scrambling the spectral
  form factor: Unitarity constraints and exact results}},\ }\href
  {https://doi.org/10.1103/PhysRevD.95.126008} {\bibfield  {journal} {\bibinfo
  {journal} {Phys. Rev. D}\ }\textbf {\bibinfo {volume} {95}},\ \bibinfo
  {pages} {126008} (\bibinfo {year} {2017})}\BibitemShut {NoStop}%
\bibitem [{\citenamefont {Deutsch}(1991)}]{Deutsch1991}%
  \BibitemOpen
  \bibfield  {author} {\bibinfo {author} {\bibfnamefont {J.~M.}\ \bibnamefont
  {Deutsch}},\ }\bibfield  {title} {\bibinfo {title} {{Quantum statistical
  mechanics in a closed system}},\ }\href
  {https://doi.org/10.1103/PhysRevA.43.2046} {\bibfield  {journal} {\bibinfo
  {journal} {Phys. Rev. A}\ }\textbf {\bibinfo {volume} {43}},\ \bibinfo
  {pages} {2046} (\bibinfo {year} {1991})}\BibitemShut {NoStop}%
\bibitem [{\citenamefont {Srednicki}(1994)}]{Srednicki1994}%
  \BibitemOpen
  \bibfield  {author} {\bibinfo {author} {\bibfnamefont {M.}~\bibnamefont
  {Srednicki}},\ }\bibfield  {title} {\bibinfo {title} {{Chaos and quantum
  thermalization}},\ }\href {https://doi.org/10.1103/PhysRevE.50.888}
  {\bibfield  {journal} {\bibinfo  {journal} {Phys. Rev. E}\ }\textbf {\bibinfo
  {volume} {50}},\ \bibinfo {pages} {888} (\bibinfo {year} {1994})}\BibitemShut
  {NoStop}%
\bibitem [{\citenamefont {Deutsch}(2018)}]{Deutsch2018}%
  \BibitemOpen
  \bibfield  {author} {\bibinfo {author} {\bibfnamefont {J.~M.}\ \bibnamefont
  {Deutsch}},\ }\bibfield  {title} {\bibinfo {title} {{Eigenstate
  thermalization hypothesis}},\ }\href
  {https://doi.org/10.1088/1361-6633/aac9f1} {\bibfield  {journal} {\bibinfo
  {journal} {Reports on Progress in Physics}\ }\textbf {\bibinfo {volume}
  {81}},\ \bibinfo {pages} {082001} (\bibinfo {year} {2018})}\BibitemShut
  {NoStop}%
\bibitem [{\citenamefont {Abanin}\ \emph {et~al.}(2019)\citenamefont {Abanin},
  \citenamefont {Altman}, \citenamefont {Bloch},\ and\ \citenamefont
  {Serbyn}}]{Abanin2019}%
  \BibitemOpen
  \bibfield  {author} {\bibinfo {author} {\bibfnamefont {D.~A.}\ \bibnamefont
  {Abanin}}, \bibinfo {author} {\bibfnamefont {E.}~\bibnamefont {Altman}},
  \bibinfo {author} {\bibfnamefont {I.}~\bibnamefont {Bloch}},\ and\ \bibinfo
  {author} {\bibfnamefont {M.}~\bibnamefont {Serbyn}},\ }\bibfield  {title}
  {\bibinfo {title} {{Colloquium: Many-body localization, thermalization, and
  entanglement}},\ }\href {https://doi.org/10.1103/RevModPhys.91.021001}
  {\bibfield  {journal} {\bibinfo  {journal} {Rev. Mod. Phys.}\ }\textbf
  {\bibinfo {volume} {91}},\ \bibinfo {pages} {021001} (\bibinfo {year}
  {2019})}\BibitemShut {NoStop}%
\bibitem [{\citenamefont {Serbyn}\ \emph {et~al.}(2021)\citenamefont {Serbyn},
  \citenamefont {Abanin},\ and\ \citenamefont {Papić}}]{Serbyn2021}%
  \BibitemOpen
  \bibfield  {author} {\bibinfo {author} {\bibfnamefont {M.}~\bibnamefont
  {Serbyn}}, \bibinfo {author} {\bibfnamefont {D.~A.}\ \bibnamefont {Abanin}},\
  and\ \bibinfo {author} {\bibfnamefont {Z.}~\bibnamefont {Papić}},\
  }\bibfield  {title} {\bibinfo {title} {{Quantum many-body scars and weak
  breaking of ergodicity}},\ }\href
  {https://doi.org/10.1038/s41567-021-01230-2} {\bibfield  {journal} {\bibinfo
  {journal} {Nature Physics}\ }\textbf {\bibinfo {volume} {17}},\ \bibinfo
  {pages} {675–685} (\bibinfo {year} {2021})}\BibitemShut {NoStop}%
\bibitem [{\citenamefont {Moudgalya}\ \emph {et~al.}(2022)\citenamefont
  {Moudgalya}, \citenamefont {Bernevig},\ and\ \citenamefont
  {Regnault}}]{Moudgalya2022}%
  \BibitemOpen
  \bibfield  {author} {\bibinfo {author} {\bibfnamefont {S.}~\bibnamefont
  {Moudgalya}}, \bibinfo {author} {\bibfnamefont {B.~A.}\ \bibnamefont
  {Bernevig}},\ and\ \bibinfo {author} {\bibfnamefont {N.}~\bibnamefont
  {Regnault}},\ }\bibfield  {title} {\bibinfo {title} {{Quantum many-body scars
  and Hilbert space fragmentation: a review of exact results}},\ }\href
  {https://doi.org/10.1088/1361-6633/ac73a0} {\bibfield  {journal} {\bibinfo
  {journal} {Reports on Progress in Physics}\ }\textbf {\bibinfo {volume}
  {85}},\ \bibinfo {pages} {086501} (\bibinfo {year} {2022})}\BibitemShut
  {NoStop}%
\bibitem [{\citenamefont {Pegg}(1998)}]{Pegg1998}%
  \BibitemOpen
  \bibfield  {author} {\bibinfo {author} {\bibfnamefont {D.~T.}\ \bibnamefont
  {Pegg}},\ }\bibfield  {title} {\bibinfo {title} {{Complement of the
  Hamiltonian}},\ }\href {https://doi.org/10.1103/physreva.58.4307} {\bibfield
  {journal} {\bibinfo  {journal} {Physical Review A}\ }\textbf {\bibinfo
  {volume} {58}},\ \bibinfo {pages} {4307–4313} (\bibinfo {year}
  {1998})}\BibitemShut {NoStop}%
\bibitem [{\citenamefont {Muga}\ \emph {et~al.}(2008)\citenamefont {Muga},
  \citenamefont {Mayato},\ and\ \citenamefont {Egusquiza}}]{Muga2008}%
  \BibitemOpen
  \bibinfo {editor} {\bibfnamefont {J.~G.}\ \bibnamefont {Muga}}, \bibinfo
  {editor} {\bibfnamefont {R.~S.}\ \bibnamefont {Mayato}},\ and\ \bibinfo
  {editor} {\bibfnamefont {{\'{I}}.~L.}\ \bibnamefont {Egusquiza}},\ eds.,\
  \href@noop {} {\emph {\bibinfo {title} {{{T}ime in {Q}uantum {M}echanics}}}}\
  (\bibinfo  {publisher} {Springer Berlin Heidelberg},\ \bibinfo {year}
  {2008})\BibitemShut {NoStop}%
\bibitem [{\citenamefont {Dodonov}\ and\ \citenamefont
  {Dodonov}(2015)}]{Dodonov2015}%
  \BibitemOpen
  \bibfield  {author} {\bibinfo {author} {\bibfnamefont {V.~V.}\ \bibnamefont
  {Dodonov}}\ and\ \bibinfo {author} {\bibfnamefont {A.~V.}\ \bibnamefont
  {Dodonov}},\ }\bibfield  {title} {\bibinfo {title} {{Energy–time and
  frequency–time uncertainty relations: exact inequalities}},\ }\href
  {https://doi.org/10.1088/0031-8949/90/7/074049} {\bibfield  {journal}
  {\bibinfo  {journal} {Physica Scripta}\ }\textbf {\bibinfo {volume} {90}},\
  \bibinfo {pages} {074049} (\bibinfo {year} {2015})}\BibitemShut {NoStop}%
\bibitem [{\citenamefont {Hilgevoord}(2002)}]{Hilgevoord2002}%
  \BibitemOpen
  \bibfield  {author} {\bibinfo {author} {\bibfnamefont {J.}~\bibnamefont
  {Hilgevoord}},\ }\bibfield  {title} {\bibinfo {title} {{Time in quantum
  mechanics}},\ }\href@noop {} {\bibfield  {journal} {\bibinfo  {journal} {Am.
  J. Phys.}\ }\textbf {\bibinfo {volume} {70}},\ \bibinfo {pages} {301}
  (\bibinfo {year} {2002})}\BibitemShut {NoStop}%
\bibitem [{\citenamefont {Hilgevoord}(2005)}]{Hilgevoord2005}%
  \BibitemOpen
  \bibfield  {author} {\bibinfo {author} {\bibfnamefont {J.}~\bibnamefont
  {Hilgevoord}},\ }\bibfield  {title} {\bibinfo {title} {{Time in quantum
  mechanics: a story of confusion}},\ }\href@noop {} {\bibfield  {journal}
  {\bibinfo  {journal} {Stud. Hist. Philos. Sci. B Stud. Hist. Philos. Modern
  Phys.}\ }\textbf {\bibinfo {volume} {36}},\ \bibinfo {pages} {29} (\bibinfo
  {year} {2005})}\BibitemShut {NoStop}%
\bibitem [{\citenamefont {Page}\ and\ \citenamefont
  {Wootters}(1983)}]{Page1983}%
  \BibitemOpen
  \bibfield  {author} {\bibinfo {author} {\bibfnamefont {D.~N.}\ \bibnamefont
  {Page}}\ and\ \bibinfo {author} {\bibfnamefont {W.~K.}\ \bibnamefont
  {Wootters}},\ }\bibfield  {title} {\bibinfo {title} {{Evolution without
  evolution: Dynamics described by stationary observables}},\ }\href
  {https://doi.org/10.1103/physrevd.27.2885} {\bibfield  {journal} {\bibinfo
  {journal} {Physical Review D}\ }\textbf {\bibinfo {volume} {27}},\ \bibinfo
  {pages} {2885} (\bibinfo {year} {1983})}\BibitemShut {NoStop}%
\bibitem [{\citenamefont {Briggs}\ and\ \citenamefont
  {Rost}(2000)}]{Briggs2000}%
  \BibitemOpen
  \bibfield  {author} {\bibinfo {author} {\bibfnamefont {J.~S.}\ \bibnamefont
  {Briggs}}\ and\ \bibinfo {author} {\bibfnamefont {J.~M.}\ \bibnamefont
  {Rost}},\ }\bibfield  {title} {\bibinfo {title} {{Time dependence in quantum
  mechanics}},\ }\href {https://doi.org/10.1007/s100530050554} {\bibfield
  {journal} {\bibinfo  {journal} {The European Physical Journal D}\ }\textbf
  {\bibinfo {volume} {10}},\ \bibinfo {pages} {311} (\bibinfo {year}
  {2000})}\BibitemShut {NoStop}%
\bibitem [{\citenamefont {Briggs}\ and\ \citenamefont
  {Rost}(2001)}]{Briggs2001}%
  \BibitemOpen
  \bibfield  {author} {\bibinfo {author} {\bibfnamefont {J.~S.}\ \bibnamefont
  {Briggs}}\ and\ \bibinfo {author} {\bibfnamefont {J.~M.}\ \bibnamefont
  {Rost}},\ }\bibfield  {title} {\bibinfo {title} {{On the Derivation of the
  Time-Dependent Equation of {S}chr{\"o}dinger}},\ }\href
  {https://doi.org/10.1023/a:1017525227832} {\bibfield  {journal} {\bibinfo
  {journal} {Foundations of Physics}\ }\textbf {\bibinfo {volume} {31}},\
  \bibinfo {pages} {693} (\bibinfo {year} {2001})}\BibitemShut {NoStop}%
\bibitem [{\citenamefont {Briggs}\ \emph {et~al.}(2007)\citenamefont {Briggs},
  \citenamefont {Boonchui},\ and\ \citenamefont {Khemmani}}]{Briggs2007}%
  \BibitemOpen
  \bibfield  {author} {\bibinfo {author} {\bibfnamefont {J.~S.}\ \bibnamefont
  {Briggs}}, \bibinfo {author} {\bibfnamefont {S.}~\bibnamefont {Boonchui}},\
  and\ \bibinfo {author} {\bibfnamefont {S.}~\bibnamefont {Khemmani}},\
  }\bibfield  {title} {\bibinfo {title} {{The derivation of time-dependent
  {S}chr{\"o}dinger equations}},\ }\href
  {https://doi.org/10.1088/1751-8113/40/6/007} {\bibfield  {journal} {\bibinfo
  {journal} {Journal of Physics A: Mathematical and Theoretical}\ }\textbf
  {\bibinfo {volume} {40}},\ \bibinfo {pages} {1289} (\bibinfo {year}
  {2007})}\BibitemShut {NoStop}%
\bibitem [{\citenamefont {Braun}\ \emph {et~al.}(2004)\citenamefont {Braun},
  \citenamefont {Strunz},\ and\ \citenamefont {Briggs}}]{Braun2004}%
  \BibitemOpen
  \bibfield  {author} {\bibinfo {author} {\bibfnamefont {L.}~\bibnamefont
  {Braun}}, \bibinfo {author} {\bibfnamefont {W.~T.}\ \bibnamefont {Strunz}},\
  and\ \bibinfo {author} {\bibfnamefont {J.~S.}\ \bibnamefont {Briggs}},\
  }\bibfield  {title} {\bibinfo {title} {{Classical limit of the interaction of
  a quantum system with the electromagnetic field}},\ }\bibfield  {journal}
  {\bibinfo  {journal} {Phys. Rev. A}\ }\textbf {\bibinfo {volume} {70}},\
  \href {https://doi.org/10.1103/physreva.70.033814}
  {10.1103/physreva.70.033814} (\bibinfo {year} {2004})\BibitemShut {NoStop}%
\bibitem [{\citenamefont {Briggs}(2015)}]{Briggs2015}%
  \BibitemOpen
  \bibfield  {author} {\bibinfo {author} {\bibfnamefont {J.~S.}\ \bibnamefont
  {Briggs}},\ }\bibfield  {title} {\bibinfo {title} {{Equivalent emergence of
  time dependence in classical and quantum mechanics}},\ }\href
  {https://doi.org/10.1103/PhysRevA.91.052119} {\bibfield  {journal} {\bibinfo
  {journal} {Phys. Rev. A}\ }\textbf {\bibinfo {volume} {91}},\ \bibinfo
  {pages} {052119} (\bibinfo {year} {2015})}\BibitemShut {NoStop}%
\bibitem [{\citenamefont {Schild}(2018)}]{Schild2018}%
  \BibitemOpen
  \bibfield  {author} {\bibinfo {author} {\bibfnamefont {A.}~\bibnamefont
  {Schild}},\ }\bibfield  {title} {\bibinfo {title} {{Time in quantum
  mechanics: A fresh look at the continuity equation}},\ }\href
  {https://doi.org/10.1103/PhysRevA.98.052113} {\bibfield  {journal} {\bibinfo
  {journal} {Phys. Rev. A}\ }\textbf {\bibinfo {volume} {98}},\ \bibinfo
  {pages} {052113} (\bibinfo {year} {2018})}\BibitemShut {NoStop}%
\bibitem [{\citenamefont {Gemsheim}\ and\ \citenamefont
  {Rost}(2023)}]{Gemsheim2023a}%
  \BibitemOpen
  \bibfield  {author} {\bibinfo {author} {\bibfnamefont {S.}~\bibnamefont
  {Gemsheim}}\ and\ \bibinfo {author} {\bibfnamefont {J.~M.}\ \bibnamefont
  {Rost}},\ }\bibfield  {title} {\bibinfo {title} {{Emergence of Time from
  Quantum Interaction with the Environment}},\ }\bibfield  {journal} {\bibinfo
  {journal} {Physical Review Letters}\ }\textbf {\bibinfo {volume} {131}},\
  \href {https://doi.org/10.1103/physrevlett.131.140202}
  {10.1103/physrevlett.131.140202} (\bibinfo {year} {2023})\BibitemShut
  {NoStop}%
\bibitem [{\citenamefont {Gemsheim}(2023)}]{Gemsheim2023phd}%
  \BibitemOpen
  \bibfield  {author} {\bibinfo {author} {\bibfnamefont {S.}~\bibnamefont
  {Gemsheim}},\ }\emph {\bibinfo {title} {{The emergence of time with
  interactions in quantum and classical mechanics}}},\ \href@noop {} {Ph.D.
  thesis},\ \bibinfo  {school} {Technische Universit{\"a}t Dresden} (\bibinfo
  {year} {2023})\BibitemShut {NoStop}%
\bibitem [{\citenamefont {Bender}\ and\ \citenamefont
  {Orszag}(1999)}]{Bender1999}%
  \BibitemOpen
  \bibfield  {author} {\bibinfo {author} {\bibfnamefont {C.~M.}\ \bibnamefont
  {Bender}}\ and\ \bibinfo {author} {\bibfnamefont {S.~A.}\ \bibnamefont
  {Orszag}},\ }\href {https://doi.org/10.1007/978-1-4757-3069-2} {\emph
  {\bibinfo {title} {{Advanced Mathematical Methods for Scientists and
  Engineers I}}}}\ (\bibinfo  {publisher} {Springer New York},\ \bibinfo {year}
  {1999})\BibitemShut {NoStop}%
\bibitem [{\citenamefont {Picasso}\ \emph {et~al.}(2022)\citenamefont
  {Picasso}, \citenamefont {Bracci},\ and\ \citenamefont
  {d’Emilio}}]{Picasso2022}%
  \BibitemOpen
  \bibfield  {author} {\bibinfo {author} {\bibfnamefont {L.~E.}\ \bibnamefont
  {Picasso}}, \bibinfo {author} {\bibfnamefont {L.}~\bibnamefont {Bracci}},\
  and\ \bibinfo {author} {\bibfnamefont {E.}~\bibnamefont {d’Emilio}},\
  }\bibinfo {title} {{Perturbation Theory in Quantum Mechanics}},\ in\ \href
  {https://doi.org/10.1007/978-3-642-27737-5_402-4} {\emph {\bibinfo
  {booktitle} {Encyclopedia of Complexity and Systems Science}}}\ (\bibinfo
  {publisher} {Springer Berlin Heidelberg},\ \bibinfo {year} {2022})\ p.\
  \bibinfo {pages} {1–33}\BibitemShut {NoStop}%
\bibitem [{\citenamefont {Ballentine}(2014)}]{Ballentine2014}%
  \BibitemOpen
  \bibfield  {author} {\bibinfo {author} {\bibfnamefont {L.~E.}\ \bibnamefont
  {Ballentine}},\ }\href {https://doi.org/9814578576} {\emph {\bibinfo {title}
  {{Quantum Mechanics: A Modern Development}}}},\ \bibinfo {edition} {2nd}\
  ed.\ (\bibinfo  {publisher} {World Scientific},\ \bibinfo {year}
  {2014})\BibitemShut {NoStop}%
\bibitem [{\citenamefont {Wootters}(1984)}]{Wootters1984}%
  \BibitemOpen
  \bibfield  {author} {\bibinfo {author} {\bibfnamefont {W.~K.}\ \bibnamefont
  {Wootters}},\ }\bibfield  {title} {\bibinfo {title} {{"Time" replaced by
  quantum correlations}},\ }\href {https://doi.org/10.1007/bf02214098}
  {\bibfield  {journal} {\bibinfo  {journal} {International Journal of
  Theoretical Physics}\ }\textbf {\bibinfo {volume} {23}},\ \bibinfo {pages}
  {701} (\bibinfo {year} {1984})}\BibitemShut {NoStop}%
\bibitem [{Note1()}]{Note1}%
  \BibitemOpen
  \bibinfo {note} {We neglect identity operations on subsystems, unless it is
  useful to state them explicitly.}\BibitemShut {Stop}%
\bibitem [{\citenamefont {Sakurai}\ and\ \citenamefont
  {Napolitano}(2017)}]{Sakurai2017}%
  \BibitemOpen
  \bibfield  {author} {\bibinfo {author} {\bibfnamefont {J.~J.}\ \bibnamefont
  {Sakurai}}\ and\ \bibinfo {author} {\bibfnamefont {J.}~\bibnamefont
  {Napolitano}},\ }\href {https://doi.org/10.1017/9781108499996} {\emph
  {\bibinfo {title} {Modern Quantum Mechanics}}},\ \bibinfo {edition} {2nd}\
  ed.\ (\bibinfo  {publisher} {Cambridge University Press},\ \bibinfo {year}
  {2017})\BibitemShut {NoStop}%
\bibitem [{\citenamefont {Polizzi}(2009)}]{Polizzi2009}%
  \BibitemOpen
  \bibfield  {author} {\bibinfo {author} {\bibfnamefont {E.}~\bibnamefont
  {Polizzi}},\ }\bibfield  {title} {\bibinfo {title} {{Density-matrix-based
  algorithm for solving eigenvalue problems}},\ }\href
  {https://doi.org/10.1103/PhysRevB.79.115112} {\bibfield  {journal} {\bibinfo
  {journal} {Phys. Rev. B}\ }\textbf {\bibinfo {volume} {79}},\ \bibinfo
  {pages} {115112} (\bibinfo {year} {2009})}\BibitemShut {NoStop}%
\bibitem [{\citenamefont {Sierant}\ \emph {et~al.}(2020)\citenamefont
  {Sierant}, \citenamefont {Lewenstein},\ and\ \citenamefont
  {Zakrzewski}}]{Sierant2020}%
  \BibitemOpen
  \bibfield  {author} {\bibinfo {author} {\bibfnamefont {P.}~\bibnamefont
  {Sierant}}, \bibinfo {author} {\bibfnamefont {M.}~\bibnamefont
  {Lewenstein}},\ and\ \bibinfo {author} {\bibfnamefont {J.}~\bibnamefont
  {Zakrzewski}},\ }\bibfield  {title} {\bibinfo {title} {{Polynomially Filtered
  Exact Diagonalization Approach to Many-Body Localization}},\ }\href
  {https://doi.org/10.1103/PhysRevLett.125.156601} {\bibfield  {journal}
  {\bibinfo  {journal} {Phys. Rev. Lett.}\ }\textbf {\bibinfo {volume} {125}},\
  \bibinfo {pages} {156601} (\bibinfo {year} {2020})}\BibitemShut {NoStop}%
\bibitem [{\citenamefont {Vedral}(2023)}]{Vedral2023}%
  \BibitemOpen
  \bibfield  {author} {\bibinfo {author} {\bibfnamefont {V.}~\bibnamefont
  {Vedral}},\ }\bibfield  {title} {\bibinfo {title} {{Classical Evolution
  without Evolution}},\ }\href {https://doi.org/10.3390/universe9090394}
  {\bibfield  {journal} {\bibinfo  {journal} {Universe}\ }\textbf {\bibinfo
  {volume} {9}},\ \bibinfo {pages} {394} (\bibinfo {year} {2023})}\BibitemShut
  {NoStop}%
\end{thebibliography}%

\end{document}